\documentclass[twocolumn,showpacs,preprintnumbers,amsmath,amssymb,APSl,prd,nofootinbib,superscriptaddress]{revtex4-1}

\usepackage{dcolumn}% Align table columns on decimal point
\usepackage{bm}% bold math
\usepackage{ifpdf}
\usepackage{hyperref}
\usepackage{dcolumn}
\usepackage{bm}
\usepackage[spanish,english]{babel}
\usepackage{amsfonts}
\usepackage{amssymb}
\usepackage{graphicx}
\usepackage{color}
\usepackage[latin1]{inputenc}

\newcommand{\LL}{\mathcal{L}}

\newcommand{\be}{\begin{equation}}
\newcommand{\en}{\end{equation}}
\newcommand{\bea}{\begin{eqnarray}}
\newcommand{\ena}{\end{eqnarray}}
\newcommand{\bes}{\begin{subequations}}
\newcommand{\ees}{\end{subequations}}

\begin{document}

\title{Scalar geons in Born-Infeld gravity}

\author{V. I. Afonso} \email{viafonso@df.ufcg.edu.br}
\affiliation{Unidade Acadêmica de F\'\i sica, Universidade Federal de Campina
Grande, 58109-970 Campina Grande, PB, Brazil}
\affiliation{Departamento de F\'{i}sica Te\'{o}rica and IFIC, Centro Mixto Universidad de Valencia - CSIC.
Universidad de Valencia, Burjassot-46100, Valencia, Spain}
\author{Gonzalo J. Olmo} \email{gonzalo.olmo@uv.es}
\affiliation{Departamento de F\'{i}sica Te\'{o}rica and IFIC, Centro Mixto Universidad de Valencia - CSIC.
Universidad de Valencia, Burjassot-46100, Valencia, Spain}
\affiliation{Departamento de F\'isica, Universidade Federal da
Para\'\i ba, 58051-900 Jo\~ao Pessoa, Para\'\i ba, Brazil}
\author{D. Rubiera-Garcia} \email{drgarcia@fc.ul.pt}
\affiliation{Instituto de Astrof\'{\i}sica e Ci\^{e}ncias do Espa\c{c}o, Faculdade de
Ci\^encias da Universidade de Lisboa, Edif\'{\i}cio C8, Campo Grande,
P-1749-016 Lisbon, Portugal}

\date{\today}
\begin{abstract}
The existence of static, spherically symmetric, self-gravitating scalar field solutions in the context of Born-Infeld gravity is explored. Upon a combination of analytical approximations and numerical methods, the equations for a free scalar field (without a potential term) are solved, verifying that the solutions recover the predictions of General Relativity far from the center but finding important new effects in the central regions. We find two classes of objects depending on the ratio between the Schwarzschild radius and a length scale associated to the Born-Infeld theory: massive solutions have a wormhole structure, with their throat at  $r\approx 2M$,  while for the lighter configurations the topology is Euclidean. The total energy density of these solutions exhibits a solitonic profile with a maximum peaked away from the center, and located at the throat whenever a wormhole exists. The geodesic structure and curvature invariants are analyzed for the various configurations considered.
\end{abstract}


\maketitle

\section{Introduction}

The notion of regular, self-gravitating fields in the context of gravitation has been at debate in the literature for decades, in particular grounded by the notion of \emph{geon} (gravitational-electromagnetic entity) introduced by J. A. Wheeler in 1955 \cite{Wheeler,LobosBook}. On its first version, geons were conjectured to exist as \emph{balls of light}, namely, as electromagnetic beams with such a high intensity that they would be held together for very long times due to their own gravitational self-interaction. Equipped with the additional ingredient of non-trivial topologies, Misner and Wheeler initiated a pioneering approach where classical electric charges, masses and other particle-like properties would be explained as purely geometric phenomena \cite{MW}. They went forward to coin the \emph{charge-without-charge} and \emph{mass-without-mass} mechanisms, by which charge and mass would emerge as properties resulting from a sourceless flux trapped into the non-trivial topology of a wormhole.

In a broader sense, geons can be seen as solutions representing localized and non-dispersing lumps of energy held together by their own gravitational attraction, regardless of their staticity/dynamics \cite{Louko:2004ej}, topology \cite{Olmo:2017qab}, asymptotic structure \cite{Martinon:2017uyo}, or presence/lack of horizons \cite{Olmo:2015bha}.  The notion of geon thus bears a close resemblance to that of \emph{soliton}, namely, non-perturbative configurations of field theory where non-linear and dissipative effects balance each other so as to allow for the existence of long-lived excitations of the fields (see \cite{SolBooks} for some books on the topic). Typically, solitons present a localized profile for their energy density and a non-trivial vacuum structure supporting different topological charges \cite{Topo-Sol} (though non-topological solitons exist as well, see for instance \cite{NTopo-Sol}), which provide a stabilizing mechanism, at least against weak perturbations. However, a number of theorems have been established forbidding the existence of soliton solutions in a large number of field theories \cite{TheoSol}, though several approaches to circumvent them have also been developed, such as the introduction of non-canonical kinetic terms \cite{Non-canonical}. One of those theorems makes it impossible for such solitons to be supported by free gauge fields in four spacetime dimensions. Thus, it came as a big surprise when Bartnik and McKinnon \cite{BM} found particle-like solutions in the context of a Yang-Mills theory coupled to General Relativity (GR). Consequently, the discovery that gravity might stabilize solitons triggered a great deal of research on gravitational configurations with soliton-like features --see, for instance \cite{GravSol}.

Scalar fields constitute another system that has been extensively discussed in the context of gravitating configurations (for a pioneering work, see \cite{Fisher}). In GR, most of the research in that direction has been devoted to seeking hairy black holes, namely, black holes with hair supported by scalar fields with different types of self-interactions \cite{HBS} (see also \cite{VolkovReview} for some recent reviews on such configurations). These objects represent extensions of the Kerr-Newman solution in the sense that, in addition to mass, charge, and angular momentum (the three quantities that emerge out of the uniqueness theorems and no-hair conjecture \cite{NH}), they are surrounded by a cloud of scalar matter whose precise characterization is required in order to accurately describe their properties \cite{KNhair}. In particular, its presence allows for the existence of new phenomena such as superradiance \cite{Superradiance}, which may trigger a ``black hole bomb" instability \cite{BBB}). In addition to hairy black holes, ``solitonic" configurations supported by scalar fields, such as boson stars \cite{Liebling}, gravitating skyrmions \cite{GS}, long-lived, quasi-stationary configurations around black holes \cite{NSG}, as well as other types of scalar solutions \cite{Sol-other} have also been considered. Moreover, the investigation on this kind of scalar field structures may be able to reveal deviations from GR \cite{fs}, for instance via scalar-tensor theories \cite{st}.

In the present work we shall follow a different route and explore the possibility of finding geon-type configurations in modified theories of gravity coupled to \emph{free} scalar fields. More specifically, the gravitational sector is taken to be an extension of GR dubbed Born-Infeld gravity \cite{BIg}, which has attracted a great deal of attention in the last few years due to its many applications in astrophysics, black hole physics, and cosmology \cite{BIg-app} (for a recent review, see \cite{bhor17}). Interestingly, this theory has shown the ability to resolve spacetime singularities in a number of black hole \cite{ors1} and cosmological \cite{oor14} scenarios. Regarding the former, it has been found that when Born-Infeld gravity is coupled to (static and spherically symmetric) electromagnetic fields, an explicit realization of Wheeler's geon arises \cite{OlmoRub}. This result is related to the emergence of a wormhole structure replacing the GR point-like singularity of the Reissner-Nordstr\"om solution of the Einstein-Maxwell field equations, a feature overlooked in preliminary analyses of this setting \cite{BF} and with deep physical consequences. Despite the generic existence of curvature divergences at the wormhole throat, an in-depth analysis has revealed that the above mentioned geonic geometries are geodesically complete \cite{ors15a}, that is, physical observers are not unavoidably destroyed during their transit through the divergent curvature region \cite{ors2}, and the scattering of scalar waves off the wormhole is well posed \cite{ors1}, thus representing non-singular solutions\footnote{Let us note that regular geometries supported by electromagnetic fields exist in GR under the form of a bundle of magnetic flux lines in static equilibrium held together by their own gravitational interaction, so-called Melvin Universe \cite{Melvin}. Such solutions also exist in the Born-Infeld gravity setup, see \cite{Melvin-cor}.}.

It is worth pointing out that the above results follow from formulating Born-Infeld gravity in the Palatini approach, where no a priori constraint between metric and connection is introduced (see {\it e.g.} \cite{Zanelli} for a pedagogical discussion). This is needed in order to ensure the second-order and ghost-free character of the resulting field equations, which is indeed a generic feature of the Palatini formulation of modified theories of gravity \cite{Franca}. This is opposed to the higher-order field equations that generically arise within the standard metric formulation, where the metric is forced to be compatible with the connection (see {\it e.g.} \cite{ReviewsPal} for some discussion on this issue). Following the finding of electromagnetic geons described above in the context of Born-Infeld gravity, the main aim of this work is to investigate the potential existence of \emph{scalar geons}, namely, static, spherically symmetric scalar fields \emph{without} a potential term, in order to stick ourselves to the original spirit of geons as free, regular, self-gravitating configurations. This research is further motivated by the possibility of finding (horizonless) compact objects as alternatives to black holes, which could have an impact on the study of gravitational waves echoes \cite{GWe,GWeb} following the observational results of LIGO \cite{LIGO}, or as sources of dark matter different from fundamental particles \cite{Lobo:2013prg,or13}. We will indeed show that, depending on a typical scale determined by the interplay between the Schwarzschild mass and the Born-Infeld parameter, two classes of configurations are found: the massive ones have a wormhole structure with a minimum nonzero area, while for the lighter ones the radial coordinate extends all the way down to $r=0$. Nonetheless, in both cases a solitonic energy density profile is found, resulting from the computation of total energy made up of the gravitational and scalar field contributions. The implications of the wormhole/non-wormhole structure for the regularity of the corresponding objects will be investigated making use of both curvature scalar and the completeness of geodesics.

This paper is organized as follows: in Sec.\ref{sec:II} we introduce the action and conventions, obtain the field equations and cast them into a suitable form for their resolution in Sec.\ref{sec:III}. Analytical approximations in the limits of interest are performed in Sec.\ref{sec:IV} to understand the structure of the field equations, while in Sec.\ref{sec:V} numerical methods are used to solve them. The solitonic character of the solutions is analyzed in Sec.\ref{sec:VI}, while a different branch of solutions is characterized in Sec.\ref{sec:VII-B}. Gathering all these results, in Sec.\ref{sec:VII} we discuss in detail the geodesic structure of both branches of solutions. We conclude in Sec.\ref{sec:VIII} with a summary and a discussion on the interpretation of the results found.

\section{Born-Infeld gravity coupled to a scalar field} \label{sec:II}

\subsection{Action and basic field equations}

The action of the Born-Infeld gravity theory with matter can be written as

\begin{eqnarray} \label{eq:action}
S &=& \frac{1}{\epsilon\kappa^2} \int d^4x \left[ \sqrt{-|g_{\mu\nu}+\epsilon R_{(\mu\nu)}| } - \lambda \sqrt{-g } \right] \nonumber \\
&&+ S_m(g_{\mu\nu},\psi_m),
\end{eqnarray}
with the following definitions and conventions: $\kappa^2 \equiv 8\pi G/c^4$ is Einstein's constant; $\epsilon$ is a parameter with dimensions of length squared; $g$ is the determinant of the spacetime metric $g_{\mu\nu}$ (and vertical bars will also denote determinants); the Ricci tensor (parenthesis denote the symmetric part) is defined from the Riemman tensor as $R_{\mu\nu}(\Gamma) \equiv {R^\rho}_{\mu\rho\nu}(\Gamma)$ where ${R^\alpha}_{\beta\mu\nu}(\Gamma)=\partial_{\mu}\Gamma^{\alpha}_{\nu\beta}-\partial_{\nu}\Gamma^{\alpha}_{\mu\beta}+\Gamma^{\alpha}_{\mu\lambda}\Gamma^{\lambda}_{\nu\beta}
-\Gamma^{\alpha}_{\nu\lambda}\Gamma^{\lambda}_{\mu\beta}$ is a function solely of the affine connection $\Gamma_{\mu\nu}^{\lambda}$ (assumed symmetric for simplicity\footnote{See \cite{Torsion} for a discussion of the field equations in Palatini theories of gravity with torsion.}), which is {\it a priori} independent of the metric $g_{\mu\nu}$ (Palatini or metric-affine formalism); the parameter $\lambda$ is related to the cosmological constant as $\lambda=1+\epsilon \Lambda_{eff}$ (this follows from expansion of the action (\ref{eq:action}) in series of $|\epsilon| \ll 1$), and $S_m$ is the matter action with $\psi_m$ denoting collectively the matter fields. For notational convenience, we will denote the object inside the first square root as $q_{\mu\nu}\equiv g_{\mu\nu} + \epsilon R_{(\mu\nu)}(\Gamma)$, which is symmetric by construction.

Variation of the action (\ref{eq:action}) with respect to metric and connection leads to the two systems of equations

\begin{eqnarray}
\sqrt{-q} q^{\mu\nu}- \sqrt{-g} \lambda g^{\mu\nu}&=&-\kappa^2 \epsilon \sqrt{-g} T^{\mu\nu}\,,  \label{eq:metric} \\
\nabla_{\alpha} \left(\sqrt{-q} q^{\mu\nu} \right)&=&0. \label{eq:connection}
\end{eqnarray}
where $q$ is the determinant of the \emph{auxiliary} metric $q_{\mu\nu}$, $T^{\mu\nu}\equiv \frac{2}{\sqrt{-g}} \frac{\delta S_m}{\delta g_{\mu\nu}}$ is the stress-energy tensor of the matter, and covariant derivatives are taken with respect to the independent connection $\Gamma_{\mu\nu}^{\lambda}$. To solve these equations one first notes that Eq.(\ref{eq:connection}), which is fully equivalent to $\nabla_{\alpha}\,q_{\mu\nu}=0$, simply expresses the compatibility between the independent connection $\Gamma_{\mu\nu}^{\lambda}$ and the metric $q_{\mu\nu}$, {\it i.e.} the former is given by the Christoffel symbols of the latter, that is

\begin{equation}
\Gamma^\lambda_{\mu\nu}= \frac{1}{2}q^{\lambda\alpha}\left(\partial_\mu q_{\alpha\nu}+\partial_\nu q_{\alpha\mu}-\partial_\alpha q_{\mu\nu}\right) \ .
\end{equation}
From the definition $q_{\mu\nu}\equiv g_{\mu\nu} + \epsilon R_{(\mu\nu)}(\Gamma)$, one can introduce a {\it deformation matrix}  $\,{\Omega^\alpha}_\nu$ by writing

\begin{equation} \label{eq:qg}
q_{\mu\nu} =  g_{\mu\alpha}{\Omega^\alpha}_\nu \ ,
\end{equation}
which, according to  (\ref{eq:metric}), satisfies the algebraic equation

\begin{equation} \label{eq:veromega}
|\Omega|^{1/2}{(\Omega^{-1})^\mu}_\nu=\lambda {\delta^{\mu}}_{\nu}  -\epsilon\kappa^2{T^\mu}_\nu \ ,
\end{equation}
where $|\Omega| \equiv \det{\Omega^\mu}_{\nu}$, for a more compact notation.
It is important to realize that solving this equation provides a relation ${\Omega^\mu}_{\nu}={\Omega^\mu}_{\nu}({T^\alpha}_{\beta})$, which means that the deformation matrix ${\Omega^\alpha}_{\nu}$ that relates the two metrics in Eq.(\ref{eq:qg}) can be solely expressed as a function of the matter sources and the metric $g_{\mu\nu}$.

Now, to write the field equations (\ref{eq:metric}) in amenable form for calculations, we contract them with the metric $g_{\alpha\nu}$ and, using Eq.(\ref{eq:qg}), we obtain the result
\begin{equation} \label{eq:Rmunuq}
{R^\mu}_{\nu}(q)=\frac{\kappa^2}{| \Omega |^{1/2}} \left(\mathcal{L}_G {\delta^\mu}_{\nu} + {T^\mu}_{\nu} \right) \ ,
\end{equation}
where ${R^\mu}_{\nu}(q) \equiv q^{\mu\alpha}R_{(\alpha \nu)}$, and $\mathcal{L}_G$ denotes the Born-Infeld gravity Lagrangian, which can be written as

\begin{equation}
\mathcal{L}_G=\frac{ | \Omega |^{1/2} - \lambda}{\epsilon \kappa^2} \ .
\end{equation}
The system of equations (\ref{eq:Rmunuq}) has several appealing features. First, as all the terms on the right-hand-side are just functions of the matter fields, they represent a system of  Einstein-like second-order field equations for the metric $q_{\mu\nu}$, with all the sources representing a modified stress-energy tensor. Since the physical metric $g_{\mu\nu}$ is related to the auxiliary one $q_{\mu\nu}$ via the matter-dependent matrix ${\Omega^\alpha}_{\nu}$ appearing in the transformation (\ref{eq:qg}), the field equations for $g_{\mu\nu}$ will be second-order as well. Second, one finds that in vacuum, ${T^\mu}_{\nu}=0$, the field equations (\ref{eq:Rmunuq}) recover GR plus a cosmological constant term, which implies the absence of extra propagating degrees of freedom. These two properties seem to be a generic feature of the Palatini formulation of classical theories of gravity \cite{Jimenez:2015caa,Bazeia:2015zpa,Franca}.

\subsection{Scalar matter}

As the matter sector of our theory we take a scalar field described by the standard action

\begin{equation} \label{eq:matteraction}
\mathcal{S}_m=-\frac{1}{2}\int d^4x \sqrt{-g} \LL_m=-\frac{1}{2}\int d^4x \sqrt{-g} (X+2V(\phi))
\end{equation}
where $X\equiv g^{\mu\nu}\partial_\mu\phi\partial_\nu\phi$ is the kinetic term and $V(\phi)$ the potential. In this work we shall assume a static spherically symmetric spacetime, whose line element can be conveniently written as

\begin{equation} \label{eq:metricg}
ds^2=-A(x)dt^2+r^2(x)d\Omega^2+\frac{1}{B(x)}dx^2 \ ,
\end{equation}
where $d \Omega^2=d \theta^2 + \sin^2(\theta) d\varphi^2$ is the angular sector, while $A(x)$, $B(x)$ and $r(x)$ are three independent functions to be determined via the gravitational plus matter field equations. The latter follow from variation of the matter action (\ref{eq:matteraction}) with respect to the scalar field $\phi$, which yields

\begin{equation} \label{eq:phi}
\Box \phi-V_\phi= \frac{1}{r^2\sqrt{A/B}}\partial_x\left(r^2\sqrt{A B}\phi_x\right)-V_\phi=0 \ ,
\end{equation}
with the notation  $\phi_x \equiv \frac{d\phi}{dx}$ and $V_\phi \equiv \frac{dV}{d\phi}$, while the stress-energy tensor reads

\begin{equation} \label{eq:tmunus}
{T^\mu}_{\nu}=g^{\mu\alpha}\partial_{\alpha}\phi\partial_{\nu}\phi-\frac{\LL_m}{2}{\delta^\mu}_{\nu} = \left(\begin{array}{lr}  -\frac{\LL_m}{2} I_{3\times3} & 0 \\ 0 & B\phi_x^2-\frac{\LL_m}{2} \end{array}\right) \ .
\end{equation}
The identity $I_{3\times 3}$ matrix represents the $(t,\theta,\varphi)$ sector, while the remaining component of the stress-energy tensor in (\ref{eq:tmunus}) can be written as ${T^x}_x=B\phi_x^2-\frac{\LL_m}{2}=\frac{1}{2} B\phi_x^2-V$. Consistently with the structure of the stress-energy tensor (\ref{eq:tmunus}) we assume for ${\Omega^\mu}_\nu$ an ansatz of the form
\begin{equation}
{\Omega^\mu}_\nu=\left(\begin{array}{lr}  \Omega_+ I_{3\times3} & 0 \\ 0 & \Omega_- \end{array}\right) \ ,
\end{equation}
From the definition (\ref{eq:veromega}) and the expression (\ref{eq:tmunus}) it follows that the two components of this ansatz read

\begin{eqnarray}
\Omega_+\!&=& \!\left(\lambda+\epsilon \kappa^2V+\tfrac{\epsilon\kappa^2}{2}B\phi_x^2\right)^{1/2}\!\!
\left(\lambda+\epsilon \kappa^2V-\tfrac{\epsilon\kappa^2}{2}B\phi_x^2\right)^{1/2} \;\label{eq:omp}\\
{\Omega_-}\!&=&\! {\left(\lambda+\epsilon \kappa^2V+\tfrac{\epsilon\kappa^2}{2}B\phi_x^2\right)^{3/2}}\!
{\left(\lambda+\epsilon \kappa^2V-\tfrac{\epsilon\kappa^2}{2}B\phi_x^2\right)^{1/2}}\;\; \label{eq:omm} \ .
\end{eqnarray}
Inserting these results into Eq.(\ref{eq:Rmunuq}), the field equations for this scalar matter source become:

\begin{widetext}
\begin{equation} \label{eq:BIscalar}
\epsilon {R^\mu}_\nu (q)=\frac{1}{\sqrt{|\Omega|}}\left(\begin{array}{lr}  \left(\sqrt{|\Omega|}-\lambda-\frac{\epsilon\kappa^2}{2}\left[B\phi_x^2+2V\right]\right)I_{3\times 3} & 0 \\ 0 & \sqrt{|\Omega|}-\lambda+\frac{\epsilon\kappa^2}{2}(B\phi_x^2 -2V) \end{array}\right) \ ,
\end{equation}
\end{widetext}
where $|\Omega|=\Omega_{+}^3\Omega_{-}$.

To further specify our setup we recall that we are looking for geonic solutions, {\it i.e.}, self-gravitating \emph{free} fields, which means that they are not supported by any degeneracy on the vacuum solutions of the potential (like, for instance, in the case of solitonic solutions supported by topologically non-trivial configurations in a flat spacetime \cite{Topo-Sol}), and thus we take $V(\phi)=0$. This way the scalar field equation (\ref{eq:phi}) can be simply integrated as

\begin{equation} \label{eq:scalarfield}
r^2\sqrt{AB}\phi_x=C \ ,
\end{equation}
where $C$ is an integration constant. Therefore, the form of the scalar field is completely specified by the metric functions, {\it i.e.}, the geometry determines the form of the scalar field. We thus find that the function $X=B\phi_x^2$ that appears in the scalar field Lagrangian $\LL_m$ and on its field equations can be written as $X=\frac{C^2}{r^4 A}$, which has an interesting formal similarity with the solution corresponding to an electric field found in \cite{OlmoRub}, where $F_{\mu\nu}F^{\mu\nu}\propto \frac{q^2}{r^4}$ for a spherical charge distribution. With the above assumptions and defining $X_\epsilon\equiv \frac{\epsilon \kappa^2}{2} X = \frac{\epsilon \kappa^2 C^2}{2\,r^4A}$, then Eqs.(\ref{eq:omp}) and (\ref{eq:omm}) become

\begin{equation} \label{eq:omegas}
\Omega_+ = {(\lambda^2 - X_\epsilon^2)^{\frac12}}
\hspace{0.1cm};\hspace{0.2cm}
\Omega_- = {(\lambda + X_\epsilon)^{\frac32}}{(\lambda - X_\epsilon)^{-\frac12}} \ ,
\end{equation}
and the field equations (\ref{eq:BIscalar}) can be written under the compact form

\begin{equation} \label{eq:RTBI}
\epsilon {R^\mu}_\nu (q)=\left(\begin{array}{lr}  \left(1-\frac{(\lambda+X_\epsilon)}{\sqrt{|\Omega|}}\right)I_{3\times 3} & 0 \\ 0 & 1-\frac{(\lambda-X_\epsilon)}{\sqrt{|\Omega|}} \end{array}\right) \ ,
\end{equation}
with $|\Omega|=\Omega_{+}^3\Omega_{-}= (\lambda + X_\epsilon)^{3} (\lambda - X_\epsilon) $.
The equations are now ready for working out their solutions.

\section{Explicit form of the field equations} \label{sec:III}

To solve the field equations (\ref{eq:RTBI}), in this section we shall closely follow Wyman's approach \cite{Wyman}, corresponding to the problem of a gravitating scalar field in GR (see also \cite{Janis:1968zz} for previous relevant results on self-gravitating, free scalar fields). On this approach one employs the scalar field as a radial coordinate, {\it i.e.},  we take $\phi_x$ to be a constant, $\phi_x=v_0$. This allows to write the following two line elements for the spacetime and auxiliary metrics, suitably adapted to our problem (see Appendix \ref{sec:App1} for full details on the justification of this choice):
\begin{eqnarray}
d{s}^2&=&-e^{{\nu}}dt^2+\frac{1}{C_0^2 {W}^4 e^{-\nu}}dx^2+\frac{1}{{W}^2}d\Omega^2 \ ,\label{ds}\\
d\tilde{s}^2&=&-e^{\tilde{\nu}}dt^2+\frac{1}{C_0^2 \tilde{W}^4 e^{-\tilde{\nu}}}dy^2+\frac{1}{\tilde{W}^2}d\Omega^2 \  ,\label{dst}
\end{eqnarray}
where the constant $C_0=C/v_0$ while the relation between the  metric coefficients $\nu, \tilde{\nu}, W, \tilde{W}$, and the radial coordinates $x$ and $y$ follows from the transformations (\ref{eq:qg}) as

\begin{eqnarray}
e^{\tilde{\nu}}&=&\Omega_+e^{\nu} \hspace{0.1cm};\hspace{0.5cm}
\tilde{W}^2=W^2/\Omega_+ \label{change} \\
dy&=& \frac{\Omega_-}{\vert \Omega \vert^{1/2}} dx= |\lambda-X_\epsilon|^{-1} dx \ . \label{eq:dydx}
\end{eqnarray}
With the above line elements, one readily verifies that the scalar field equation is just $\phi_{xx}=0$, which implies $\phi=v_0 x +\phi_0$, where $\phi_0$ is an integration constant. Now, the components of the Ricci tensor for the metric $q_{\mu\nu}$ in Eq.(\ref{ds}) become

\begin{eqnarray}
{R^t}_t &=& -\tfrac{1}{2} C_0^2\tilde{W}^4 e^{-\tilde{\nu}} \tilde{\nu}_{yy}\\
{R^y}_y &=& -\tfrac{1}{2} C_0^2\tilde{W}^3 e^{-\tilde{\nu}} \left(4\tilde{W}_y\tilde{\nu}_y-4\tilde{W}_{yy}+\tilde{W}\tilde{\nu}_{yy}\right) \\
{R^\theta}_\theta &=& \tilde{W}^2\left(1+C_0^2e^{-\tilde{\nu}}[\tilde{W}\tilde{W}_{yy}-\tilde{W}_y^2]\right) \ ,
\end{eqnarray}
which are needed to solve the field equations (\ref{eq:RTBI}). Now, since the right-hand side of such equations can be read as a modified stress-energy tensor ${\tau^\mu}_{\nu}$, from the combinations  ${R^t}_t={\tau^t}_t$, ${R^y}_y-{R^t}_t={\tau^y}_y-{\tau^t}_t$, and ${R^\theta}_\theta={\tau^\theta}_\theta$,
the field equations (\ref{eq:RTBI}) can be expressed as

\begin{eqnarray}
\tilde{\nu}_{yy}&=&-\frac{\kappa_0^2 \Omega_+^3}{X_\epsilon}\left(1-\tfrac{\lambda+X_\epsilon}{\Omega^{1/2}}\right)  \label{fulleq1}\\
0&=& \tilde{W}_{yy}-\tilde{\nu}_y\tilde{W}_y- \frac{ \kappa_0^2\Omega_+^3}{2\Omega^{1/2}}\tilde{W} \label{fulleq2}\\
\tilde{W}\tilde{W}_{yy}-\tilde{W^2_y}&=& -\frac{e^{\tilde{\nu}}}{C_0^2}+ \frac{\kappa_0^2 \Omega_+^3}{2X_\epsilon} \left(1-\tfrac{\lambda+X_\epsilon}{\Omega^{1/2}}\right)\tilde{W}^2  \  ,   \label{fulleq3}
\end{eqnarray}
where $\kappa_0^2\equiv \kappa^2 v_0^2$ has dimensions of length$^{-2}$, while $C_0^2$ has dimensions of length$^4$. Our problem thus boils down to solving these equations for $\tilde{\nu}$ and $\tilde{W}$ and then use the relation between $g_{\mu\nu}$ and $q_{\mu\nu}$ to completely determine the line element generated by the scalar field. It should be noted that Eq.(\ref{fulleq3}) is a constraint that the solutions to (\ref{fulleq1}) and (\ref{fulleq2}) must satisfy. As will be seen later, this constraint will introduce a relation between the two integration constants that characterize the function $\tilde{W}$ [see Eq.(\ref{eq:constraintGR}), (\ref{eq:Winf}) and (\ref{eq:Wep2})] .

Given the strongly nonlinear character of Eqs.(\ref{fulleq1}), (\ref{fulleq2}) and (\ref{fulleq3}), numerical methods will be necessary to find solutions. It must be noted, however, that there is still an algebraic part we must take care of before being able to integrate these equations numerically. The difficulty lies on the fact that the variable $X_\epsilon=(\epsilon \kappa_0^2/2) C^2 W^4 e^{-\nu}$ is referred to the variables $W$ and $\nu$, while our equations involve derivatives of $\tilde{W}$ and $\tilde{\nu}$. In order to rewrite $X_\epsilon$ in terms of the variables $\tilde{W}$ and $\tilde{\nu}$, let us define $\theta \equiv (\epsilon \kappa_0^2/2) C^2\tilde{W}^4 e^{-\tilde{\nu}}$ so we have
\begin{equation}
X_\epsilon=\frac{\epsilon \kappa_0^2}{2} C^2 W^4 e^{-\nu}= \Omega_+^3\theta =(\lambda^2-X_\epsilon^2)^{3/2}\,\theta\ .
\end{equation}
The square of this equation can be straightforwardly solved and leads to the relation
\begin{equation}\label{eq:solX2}
X_\epsilon^2(\theta) =\lambda^2 + \frac {1} {18^{1/3} |\theta|} [K (\theta)- 12^{1/3}/ K(\theta)]\,,
\end{equation}
with $K(\theta)=\left(-(9\lambda^2|\theta|) + \sqrt {12 + (9\lambda^2|\theta|)^2}\right)^{1/3}$.

By definition, $X_\epsilon$ is linear in $\epsilon$. In the following we will take the negative branch of $\epsilon$ as it directly leads to nonsingular  bouncing solutions in cosmological models \cite{oor14} and nonsingular black hole spacetimes in electrovacuum configurations \cite{ors1,ors2}. Therefore, we shall keep the negative branch of the square root in \eqref{eq:solX2}. In Sec.\ref{sec:VII-B} we will consider the $\epsilon>0$ case for completeness. The field equations are now almost ready for their analysis but, first, we need to precise a bit more the notation that will be employed hereafter.

\subsection{Comments on the notation}

Before proceeding further, it should be noted that $X_\epsilon$ is a dimensionless quantity, which can be used to introduce some useful notation in dimensionless form. In particular, introducing the length-squared scales  $\epsilon\equiv-2l_\epsilon^2$ and $r_C^2\equiv \kappa_0^2 C^2$, the fact that $W$ has dimensions of inverse length allows us to define $\hat W\equiv r_\epsilon W $ such that $r_\epsilon^4= l_\epsilon^2 r_C^2$. This turns $X_\epsilon$ into $X_\epsilon= -\hat W^4 e^{-\nu}$.  Similarly, one can write $\theta=-\hat{\tilde{W}}^4 e^{-\tilde{\nu}}$. With this notation, Eqs.(\ref{fulleq1})-(\ref{fulleq3}) can be regarded as dimensionless, with the variable $y$ replaced by $\hat y=y/r_\epsilon$, $\tilde{W}$ by $\hat{\tilde{W}}$, $\kappa_0^2$ by $\hat\kappa^2_0\equiv \kappa_0^2r_{\epsilon}^2$ (dimensionless), and $C_0^2$ by $\hat C_0^2\equiv C_0^2/r_\epsilon^4$. The line element $d\tilde{s}^2$ in (\ref{dst}) should thus be seen as the dimensionless quantity $d\tilde{s}^2/r_\epsilon^2$, with $dt\to  d\hat t$ and $dy\to d\hat y$ also dimensionless (or measured in units of $r_\epsilon$, which is the natural scale of the problem). From now on we will use this dimensionless form of the field equations but omitting the {\it hats} to avoid unnecessary redefinitions. Nonetheless, to be fully consistent, we will also use the notation $z\equiv r/r_\epsilon$ and $\tilde{z}\equiv \tilde{r}/r_\epsilon$ for the radial coordinates.

\section{Analytic approximations} \label{sec:IV}

\subsection{GR limit (asymptotic behavior, $\theta\to 0$)}\label{sec:GRlimit}

According to the redefinitions that lead from the line element (\ref{eq:ds2rt}) to (\ref{dst}) and the above notation, it is clear that $\tilde{W}= 1/\tilde{z}$. This implies that the limit $\theta\to 0$ represents the asymptotic far region, whereas $\theta\to \infty$ must be seen as the internal region (this point will be verified numerically later). When $\theta\to 0$ one gets $X_\epsilon\approx \theta=- |\theta|$ (recall that  $\theta\equiv (\epsilon \kappa_0^2/2) C^2\tilde{W}^4 e^{-\tilde{\nu}} < 0$).

With the dimensionless notation introduced in the last section we have $\theta\equiv -\tilde{W}^4 e^{-\tilde{\nu}}$. Then, in the asymptotic limit $\theta\to 0$ ($z\gg 1$ and $e^{-\nu}\to 1$), and setting $\lambda=1+\epsilon \Lambda_{eff}\to 1$ for simplicity, we find

\begin{eqnarray}
\tilde{\nu}_{yy}&\simeq &- \frac{\kappa_0^2}{2}|\theta| \\
0&\simeq&\tilde{W}_{yy}-\tilde{\nu}_y\tilde{W}_y- \frac{\kappa^2_0}{2}(1 + |\theta|)\tilde{W} \\
\tilde{W}\tilde{W}_{yy}-\tilde{W_y}^2&\simeq& -\frac{e^{\tilde{\nu}}}{C_0^2} + \frac{\kappa_0^2 }{4}|\theta| \tilde{W}^2 \,.
\end{eqnarray}
As the function $|\theta|$ assumes very small values, the GR  equations \cite{Wyman} are nicely recovered. Indeed, in that limit, the above equations become

\begin{eqnarray}
\tilde{\nu}_{yy}&=&0 \label{Wy1}\\
\tilde{W}_{yy}-\tilde{\nu}_{y}\tilde{W}_{y}- \frac{\kappa_0^2}{2} \tilde{W}&=&0  \label{Wy2}\\
\tilde{W}\tilde{W}_{yy}-\tilde{W_{y}}^2&=& - \frac{e^{\tilde{\nu}}}{C_0^2} \ , \label{Wy3}
\end{eqnarray}
which can be readily integrated as

\begin{eqnarray}
\tilde{\nu}_{Far}&=& \alpha y+\beta \label{Wynu} \\
\tilde{W}_{Far}&=&a e^{m_+ y}+ b e^{m_- y} \label{WyW1} \\
m_{\pm} &=&\tfrac{1}{2} \left(\alpha \pm \sqrt{\alpha^2 + 2\kappa_0^2}\right) \\
0&=&\frac{e^\beta}{C_0^2}+a b  (\alpha^2+2\kappa_0^2) \ . \label{eq:constraintGR}
\end{eqnarray}
where $\alpha, \beta, a, b$ are integration constants. In GR, asymptotically flat solutions require $a=-b$ \cite{Wyman} and then $a^2=e^\beta/(C_0^2 ({\alpha^2+2\kappa_0^2}))$. Since we are interested in the modifications induced by the Born-Infeld dynamics near the center,  this is the set of asymptotic boundary conditions we will use in our problem. To be more explicit and better visualize the above exact solutions, it is useful to write them in terms of the radial coordinate $\tilde{z}$. From the solution \cite{Wyman}

\begin{equation}
\tilde{W}_{Asympt.}=\frac{e^{\frac{\beta+\alpha y}{2}}}{C_0} (\sinh (\gamma y)/\gamma)\ ,
\end{equation}
where $\gamma=\sqrt{\alpha^2+2\kappa^2}/2$, the line element in the GR limit thus takes the form

\begin{eqnarray}\label{eq:GRds2}
\frac{ds^2_{GR}}{r_\epsilon^2}&=&-e^{\beta+\alpha y}dt^2+C_0^2{e^{-(\beta+\alpha y)} (\sinh\gamma y/\gamma)^{-4}}dy^2 \nonumber \\
&&+C_0^2{e^{-(\beta+\alpha y)} (\sinh\gamma y/\gamma)^{-2}}d\Omega^2 \ .
\end{eqnarray}
In the asymptotic far region ($y\to 0$ or $\tilde{z}\to \infty$), this line element turns into

\begin{equation}
\frac{d{s}^2_{GR}}{r_\epsilon^2}\approx -e^{\beta+\alpha/\tilde{z}}dt^2+\frac{C_0^2}{e^\beta}e^{-\alpha/\tilde{z}}\left(d\tilde{z}^2+\tilde{z}^2d\Omega^2\right) \ ,
\end{equation}
with $e^{\pm \alpha/\tilde{z}}\approx (1\pm\alpha/\tilde{z})$, thus confirming its asymptotic flatness. From the spatial sector of the metric, ${e^\beta}/{C_0^2}=1$ appears as a natural choice for $\beta$. The remaining $e^\beta$ term in the time component can be absorbed into a redefinition of the time coordinate. In this way, the resulting metric coincides with the far limit of the Schwarzschild solution if we take $\alpha=-2M/r_\epsilon$. Recalling the relation between $\tilde{z}$ and $\tilde{r}=r_\epsilon \tilde{z}$ and given that in the asymptotically far region $r\approx \tilde{r}$, the term $\alpha/\tilde{z}$ becomes $-2M/r$, as one would expect.

\subsection{Internal region ($|\theta|\to \infty$)}

The internal region corresponds to the limit when $|\theta|\to \infty$. Here we find that $X_\epsilon\approx -1+1/(2|\theta|^{2/3})$,  $\Omega_+\approx |\theta|^{-1/3}$, and Eqs.(\ref{fulleq1})-(\ref{fulleq3}) reduce to

\begin{eqnarray}
\tilde{\nu}_{yy}&\simeq & -\kappa_0^2 |\theta|^{-2/3} \approx 0\\
\tilde{W}_{yy}-\tilde{\nu}_y\tilde{W}_y&\simeq& \kappa_0^2(1 - \tfrac{1}{4} |\theta|^{-2/3})\tilde{W} \approx \kappa_0^2 \tilde{W} \\
\tilde{W}\tilde{W}_{yy}-\tilde{W_y}^2&\simeq& -\frac{e^{\tilde{\nu}}}{C_0^2}+\frac{ \kappa_0^2 }{2}|\theta|^{-2/3} \tilde{W}^2
 \approx -\frac{e^{\tilde{\nu}}}{C_0^2} \ . \label{eq:constr}
\end{eqnarray}
From the above limit, we see that in the central region the field equations behave exactly like in the asymptotic (GR) limit up to a redefinition of constants [compare to Eqs.(\ref{Wy1})-(\ref{Wy3})]. In this limit the field equations can also be analytically integrated yielding

\begin{eqnarray}
\tilde{\nu}_{Center}(y)&=& l_1+l_2 y  \label{eq:nuinf}\\
\tilde{W}_{Center}(y)&=& - \frac{e^{l_1}}{C_0^2 D_2 l_\kappa^2}  e^{\frac{1}{2} (l_2 - l_\kappa)  y}+D_2 e^{\frac{1}{2} (l_2 + l_\kappa) y}\,, \label{eq:Winf}\quad
\end{eqnarray}
where $l_1$ and $l_2$ are integration constants while we have defined $l_{\kappa} =\sqrt{4 \kappa_0^2+l_2^2}$ and the condition $0=\frac{e^{l_1}}{C_0^2}+D_1 D_2 l_{\kappa}^2$ that follows from  (\ref{eq:constr}) has been used. We note that given that $l_\kappa>|l_2|$, in the limit $y\to \infty$ only the second term in $\tilde{W}_{center}(y)$ survives.
As $\kappa_0$ is a fixed quantity, the numerical integration will allow us to adjust the coefficients $l_1, l_2$ and  $D_2$ once initial conditions are given in the asymptotic far region ($y\to 0$).

Before getting into the numerics, let us discuss analytically the asymptotic behavior of these solutions. It is easy to verify that in the limit $|\theta|\to \infty$, the physical line element takes the form
\begin{equation}\label{eq:ds2thetainf0}
\frac{ds^2}{r_\epsilon^2}\approx-\left(\tilde{W}^2e^{\tilde{\nu}}\right)^{2/3}dt^2+\frac{1}{C_0^2}dx^2+\left(\tilde{W}^2e^{\tilde{\nu}}\right)^{-1/3}d\Omega^2 \ ,
\end{equation}
Given that $z^2(y)\approx\left(\tilde{W}^2e^{\tilde{\nu}}\right)^{-\frac{1}{3}}$ and,  for $\epsilon<0$,  $\lim_{|\theta|\to \infty}dx^2\approx 4 dy^2$, (\ref{eq:ds2thetainf0}) can be written as
\begin{equation}\label{eq:ds2thetainf1}
\frac{ds^2}{r_\epsilon^2}\approx-\frac{1}{z^4(y)}dt^2+\frac{4}{C_0^2}dy^2+z^2(y)d\Omega^2 \ .
\end{equation}
Using the explicit relation between $z$ and $y$ specified by $z^2(y)$ and Eqs.(\ref{eq:nuinf}) and (\ref{eq:Winf}), \emph{i.e.}, $z^2=\frac{e^{-\frac{1}{3}(2l_2+l_k)y}}{D_2^{2/3}e^{l_1/3}}$, we  get
\begin{equation}
\frac{ds^2}{r_\epsilon^2}=-\frac{1}{{z}^4}dt^2+\frac{4\sigma^2}{C_0^2}\frac{d{z}^2}{{z}^2}+z^2d\Omega^2  \ ,\label{eq:ds2center}
\end{equation}
where $\sigma^2\equiv \frac{6^2}{(2l_2+l_\kappa)^2}$.
For comparison, in the GR case the line element in the $y\to \infty$ limit behaves as
\begin{eqnarray}\label{eq:ds2yGR}
\frac{ds^2_{GR}}{r_\epsilon^2}&=&-e^{\alpha y}dt^2+{16\gamma^4}e^{-2(\alpha+m_+)y}dy^2 \nonumber \\
&&+{4\gamma^2}e^{-2m_+y}d\Omega^2 \ ,
\end{eqnarray}
which is equivalent to
 \begin{equation}
\frac{ds^2_{GR}}{r_\epsilon^2}=-\left(\frac{\gamma}{z}\right)^{\frac{\alpha}{m_+}}dt^2+\left(\frac{\gamma}{z}\right)^{\frac{\alpha}{m_+}-2}\frac{\gamma^2dz^2}{m_+^2}+z^2d\Omega^2  \ .\label{eq:ds2centerGR}
\end{equation}
It is worth noting that the line element in the GR case is very sensitive to the value of $\alpha$. Since $\alpha$ is related to the asymptotic Newtonian mass as $\alpha=-2M/r_\epsilon$ (see Sec. \ref{sec:GRlimit} above) and the quotient  $\kappa_0^2/\alpha^2 \sim 0$  in physically reasonable situations\footnote{Recall that $\kappa_0^2$ actually represents the dimensionless quantity $\kappa^2 v_0^2 r_\epsilon^2$, where $\kappa^2 v_0^2$ is an inverse squared length scale associated to the amplitude of the scalar field, $\phi=v_0 x$. Given that $r_\epsilon=\sqrt{l_\epsilon r_C}$ is the natural scale of the problem, we can assume $l_\phi^2=1/\kappa^2 v_0^2$ to be bigger than $r_\epsilon^2$ or, at least, of the same order of magnitude, which implies $\kappa^2 v_0^2 r_\epsilon^2\leq1$.}, thus $\gamma =(| \alpha| /2) \sqrt{1+2\kappa_0^2/\alpha^2}
\sim |\alpha|/2$ and the ratio $\alpha/m_+ \approx -2\alpha^2/\kappa_0^2$ is expected to be negative and very large.
The behavior of  (\ref{eq:ds2center}) instead is more {\it universal},  being the exponents of the radial dependence of the metric components independent of the values of the parameters that characterize those solutions.  On the other hand, one can verify that, despite differences, the two line elements above lead to a Ricci tensor with vanishing components except for $r_\epsilon^2R_{rr}=-2(m_{-}/ m_{+}\!) \,z^{-2} ,\, R_{\theta\theta}=1, \, R_{\varphi\varphi}=\sin^2\theta$ in the GR case, and  $r_\epsilon^2R_{rr}=-6\, z^{-2}, \, R_{\theta\theta}=1, \, R_{\varphi\varphi}=\sin^2\theta$ in the Born-Infeld case, both having an $1/z^2$ dependence in this region. With additional calculations, one finds that in the Born-Infeld case the Kretschmann and Ricci scalars have a universal power-law behavior (independent of the parameters that characterize the solution) given by
\begin{eqnarray}
r_\epsilon^4{R^\alpha}_{\beta\mu\nu}{R_\alpha}^{\beta\mu\nu} &\approx & \frac{4}{z^4} -\frac{2C_0^2}{\sigma^2 z^2}+\frac{27C_0^4}{4\sigma^4} \label{eq:scalars} \\
r_\epsilon^2R&\approx&\frac{2}{z^2} -\frac{3 C_0^2}{2\sigma^2} \ ,  \nonumber \\  \nonumber
\end{eqnarray}
whereas in the GR case the power-law is very dependent on the details of the far solution
\begin{eqnarray}
r_\epsilon^4({R^\alpha}_{\beta\mu\nu}{R_\alpha}^{\beta\mu\nu})_{GR} &\approx&
\left(\frac{\gamma}{z}\right)^{-\frac{2\alpha}{m_+}} \\
&&\times \frac{4m_+^2(3m_+^2-2m_+ \alpha+2\alpha^2)}{z^8}  \nonumber \\
r_\epsilon^2 R_{GR} &\approx& -\left(\frac{\gamma}{z}\right)^{-\frac{\alpha}{m_+}}\frac{(\alpha^2+\kappa^2)}{z^4}
\end{eqnarray}
(recall that $\alpha<0$ and $m_+>0$). This puts forward that the Born-Infeld gravity dynamics has been able to soften and universalize the amplitude of curvature scalars in the interior region. Later we will study the implications of these results for the regularity of the corresponding spacetimes.

\section{Numerical analysis}\label{sec:V}

We will next find numerical solutions for the set of equations (\ref{fulleq1})-(\ref{fulleq3}) subject to the asymptotically flat initial conditions given in Sec.\ref{sec:GRlimit}. Taking advantage of the fact that in the two asymptotic regions (far and interior) an analytical expression for the solutions is known, we will also obtain functional fittings for $\tilde{\nu}$ and $\tilde{W}$. The unknown coefficients of the analytical approximations will thus be obtained by comparison with the functions fitting the numerical solutions. This way we obtain (``Fit" stands for fitting solutions)
\begin{widetext}
\begin{eqnarray}
\tilde{\nu}_{Fit}&=&C_\nu+ \left(\alpha - A_\nu \tanh(a_\nu y_c + f)\right) y
- \frac{A_\nu}{a_\nu} \log[\cosh( a_\nu (y-y_c) + f ) ] \label{eq:nufit}\\
(\tilde{W})_{Fit}&=&
8 \beta e^{\frac{1}{8\beta}y}
+\frac14 P_1 e^{\Delta_1 (y-y_c)+\Delta_2+\Delta} \left\{ \frac{12}{\Delta_1}-\frac{1}{2 \beta} \tanh\left(\frac{\Delta_1}{2\beta} y_c \right) +\right.\label{eq:Wfit}\\
&&\left. \frac{1}{1+e^{\frac{\Delta_1}{\beta} y_c}}
 \left[ \frac{e^{\frac{\Delta_1}{\beta} y}}{(1+\beta)} \, _2F_1\left(1,\beta +1\,;(\beta+1)+1\,; -e^{\frac{\Delta_1}{\beta} (y-y_c)}\right)
 - \frac{1}{ \beta} \, _2F_1\left(1,\beta\,; \beta +1\,; -e^{\frac{\Delta_1}{\beta} (y-y_c)} \right) \right]\right\} \notag\qquad
\end{eqnarray}
\end{widetext}
where
$\Delta_1=\log[\tilde{W}(y_{\infty})/\tilde{W}(y_c)]/(y_{\infty} - y_c)$,  $\Delta_2 =\log[\tilde{W}(y_c)]$, $\Delta=\log[\tilde{W}(y_{\infty})/\tilde{W}'(y_{\infty})]$ and
$\beta=1/(8\alpha)$; $y_\infty$ is the largest value assigned for the radial coordinate $y$ and the $y_c$ values correspond to the center of the energy density distribution (determined numerically) for each value of $\alpha$ (see Sec. \ref{sec:VI}). For the sake of illustration, in Table \ref{tab:I} we show the values of the fitting parameters for some representative values of $\alpha$, obtained on the support $[y_0, y_\infty] = [.001, 30]$\footnote{All numerical values and plots are based on calculations taking $\kappa=1$ and in all tables we show just four digits precision for simplicity.}

\begin{figure}[h]
\centering
\includegraphics[width=0.40\textwidth]{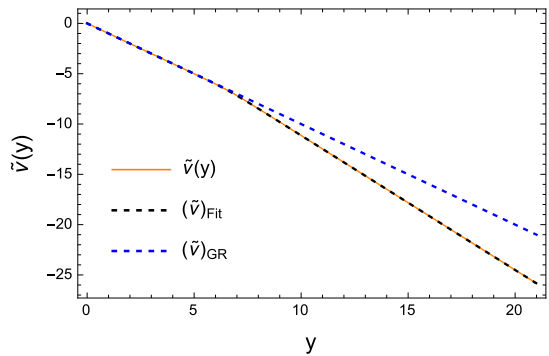}
\caption{The function $ \tilde\nu(y)$ for the numerical integration (solid orange curve), and the analytical fitting of Eq.(\ref{eq:nufit}) (dashed black) for integration constant $\alpha=-10$ and Born-Infeld length $l_{\epsilon}=10^{-3}$. The dotted blue curve represents the GR behavior ($l_{\epsilon} \rightarrow 0$), to which the curves converge as $y \rightarrow 0$. \label{fig:1}}
\end{figure}

\begin{figure}[h!]
\centering
\includegraphics[width=0.40\textwidth]{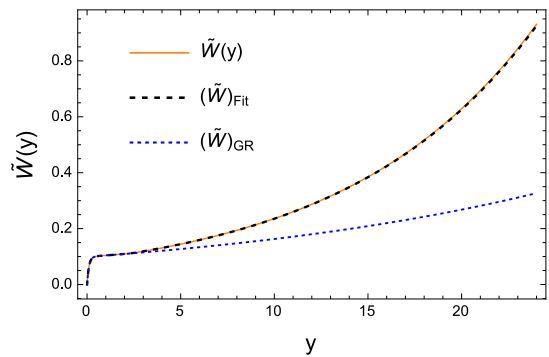}
\caption{The function $\tilde{W}(y)$ with the same notation and parameters as in Fig.~\ref{fig:1}. The analytical fitting corresponds to Eq.(\ref{eq:Wfit}). \label{fig:2}}
\end{figure}

\begin{table}[h]
\begin{center}
\begin{tabular}{c | c c c c}
$\alpha$ &-50 & -10& -1& -0.8 \\[1mm]
\hline
$y_c$ & 0.5943 & 2.2867& 6.5159 & 6.4491\label{lpfit}\\[1mm]
\hline
$C_\nu$& 0.0060 &0.1107 &1.0619&1.0542\\
$A_\nu$&0.0100&0.0483&0.1684&0.1691\\
$a_\nu$&20.7348&4.2920&1.2488&1.2437\\
$f$ &-0.4663&-0.4082&-0.2776&-0.2744\\[1mm]
\hline
$P_1$& 0.2547&0.2497&0.2666&0.2718\\
\end{tabular}
\caption{Some examples of the fitting parameters in Eqs.(\ref{eq:nufit}) and (\ref{eq:Wfit}) resulting from the numerical computations for four representative values of $\alpha$ .}
\label{tab:I}
\end{center}
\end{table}
The results of the numerical integration and the fitting functions $\tilde{\nu}_{Fit}(y)$ and $ \tilde{W}_{Fit}(y)$ are plotted in Figs. \ref{fig:1} and \ref{fig:2}, respectively, where the deviations with respect to the GR behavior for $\theta \rightarrow \infty$  ($y \rightarrow \infty $) are manifest. In such region both $\tilde{\nu}(y)$ and $\log \tilde{W}(y)$ present a linear behavior, which can be explicitly fitted to
\begin{eqnarray}
\tilde{\nu}(y)|_{y \rightarrow \infty} &=& l_1 +l_2 \,y \label{eq:nufit} \\
\log \tilde{W}(y)|_{y \rightarrow \infty} &=& l_3 + l_4 \,y \label{eq:lwfit}\\
\tilde{W}|_{y \rightarrow \infty} & \approx &D_2 \,e^{\frac12 (l_2 + l_\kappa)  y}\ , \label{eq:wfit}
\end{eqnarray}
where $l_4 = \frac{1}{2} (l_ 2 + l_{\kappa})$ with $l_{\kappa}= \sqrt{4\kappa_ 0^2 + l_ 2^2}$. The values of the unknown coefficients $l_1, l_2, l_3$ in the equations above
are numerically fitted for some representative values of $\alpha$, and displayed in Table \ref{table:II}, where we also include
the value of $D_2=\exp(l_3)$, which is directly connected to Eqs.(\ref{eq:nuinf}) and (\ref{eq:Winf}).

\begin{table}[h]
\begin{center}
\begin{tabular}{c | c c c c}
$\alpha$ &-50 & -10& -1& -0.8 \\[1mm]
\hline
$l_1$ & 0.0121 & 0.2269 & 2.2305&2.2170\\
$l_2$& -50.0200 & -10.0967 & -1.3368&-1.1382\\
\hline
$l_3$ & -3.9185 & -2.4259  & -1.6647& -1.5941\\
$D_2$ &0.0199 & 0.0884 & 0.1893&0.2031
\end{tabular}
\caption{Some examples of the values fitted numerically corresponding to the coefficients $l_1, l_2, l_3, D_2$  in Eqs.(\ref{eq:nufit}), (\ref{eq:lwfit}) and (\ref{eq:wfit}), for several values of $\alpha$. }
\label{table:II}
\end{center}
\end{table}

In terms of the above fitting solutions one can compute the metric components of \eqref{ds} and (\ref{dst}) and then analyze the scalar field configuration in the full range.

\subsection{Spherical sector} \label{sec:secSS}

From the exact GR solution presented in Eq.(\ref{eq:GRds2}) one finds that the radial function $z^2=1/W^2(y)$ approaches $z\to 0$ exponentially fast in the limit $y\to \infty$, {\it i.e.}, $z^2\approx 4C_0^2 e^{-\beta}\gamma^2 e^{-2m_+y}$, where $m_+=\frac{1}{2} (\alpha+ \sqrt{\alpha^2 + 2\kappa^2}) >0$. In the Born-Infeld gravity case under study,
we also find an exponential behavior, $z^2\approx D_2^{-2/3}e^{-l_1/3} e^{-\frac{1}{3}(2l_2+l_k)y}$, though the sign of the constants in the exponential cannot be guessed {\it a priori}. However, from the numerical analysis (see Table \ref{table:II} above), we see that the value of $l_2$ is very close to the value of the constant $\alpha$, where for each studied case we have that $l_2<0$. In fact, $l_2$ can be put in linear relation with $\alpha$, as shown in Fig. \ref{fig:l2} below.

\begin{figure}[h!]\centering
\includegraphics[width=0.45\textwidth]{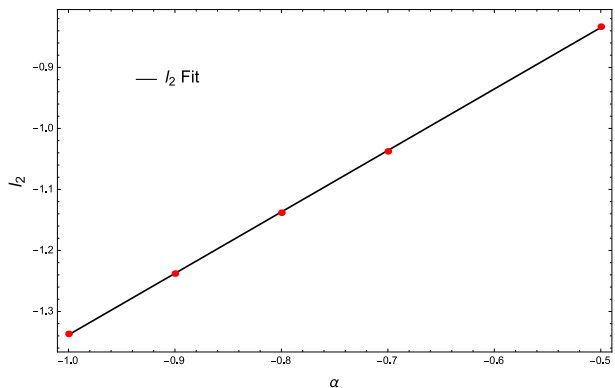}
\caption{Linear fitting of $l_2$ for the lightest values of $\alpha$ studied.}
\label{fig:l2}
\end{figure}

Now, recalling that $l_k=\sqrt{4\kappa_0^2+l_2^2}$, it is easy to see that for $l_2>-2\kappa_0/\sqrt{3}$ the combination $2l_2+l_k$ in the exponential is positive, thus implying that $z^2$ goes to zero as $y\to \infty$, like in the GR case. However, for any  $l_2<-2\kappa_0/\sqrt{3}$, the exponential changes sign and grows unboundedly as $y\to \infty$. The behavior of the solutions can thus be classified in two types depending on whether they represent massive objects, {\it i.e.}, $|l_2|\sim |\alpha|$ being large as compared to $2\kappa_0/\sqrt{3}$; or light objects, $|l_2|$ small as compared to $2\kappa_0/\sqrt{3}$. Furthermore, for massive objects the radial function $z(y)$ reaches a minimum and bounces off, which can be naturally interpreted as a signal of the presence of a wormhole\footnote{Let us stress that, as stated in the introduction, similar wormholes configurations are also found in the context of Born-Infeld gravity when electromagnetic fields are considered \cite{OlmoRub}.}, namely, a topologically non-trivial configuration connecting two regions \cite{Visser}, with that minimum representing its throat. For lighter configurations such a minimum (and thus the wormhole structure) disappears and the radial function extends all the way down to $r=0$.

\begin{figure}[h!]\centering
\includegraphics[width=8.0cm, height=7.0cm]{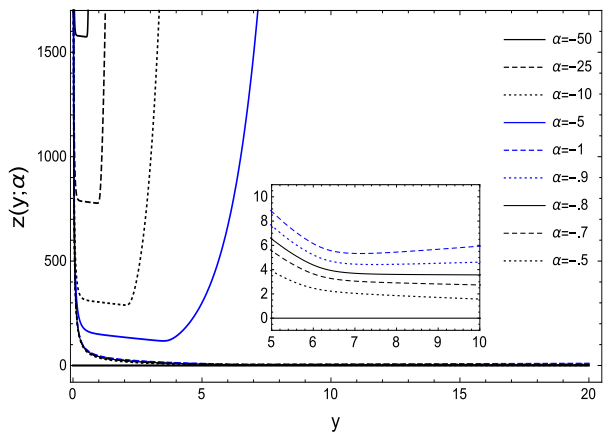}
\caption{Radial function $z=1/( \Omega_{+} \tilde{W}(y))$ (where $y=\phi^{-1}(x)$) for different values of the parameter $\alpha$, where the minimum attained by the radial function $z(y)$ is observed, corresponding to the throat of the wormhole solutions. The inset tracks the disappearance of this minimum, leading to the transition between wormhole/non-wormhole configurations.}
\label{fig:radii}
\end{figure}

\begin{figure}[h!]\centering
\includegraphics[width=0.47\textwidth]{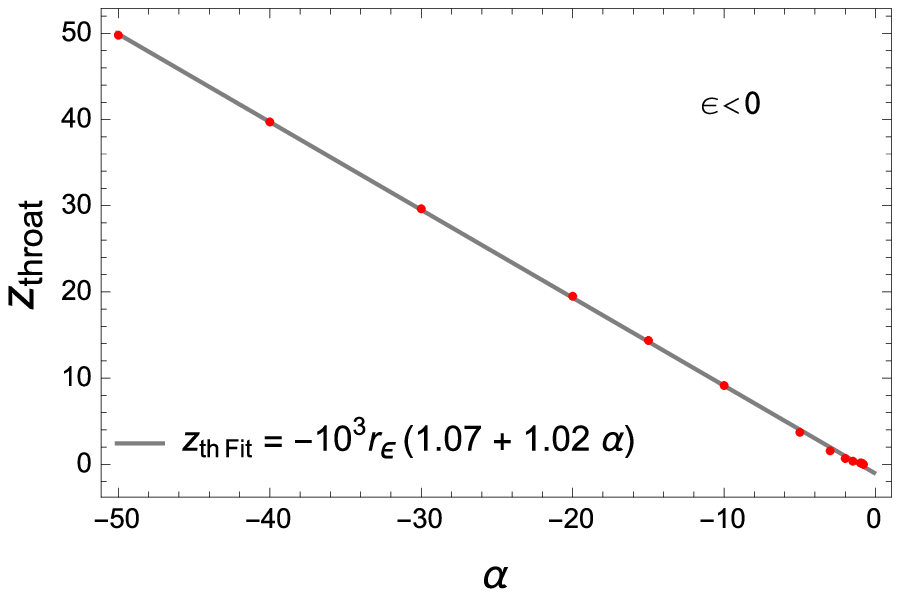}
\caption{Linear relation between the location of the wormhole throat and the parameter $|\alpha|=2M/r_\epsilon$ in the case $\epsilon<0$. }
\label{fig:rminalpha}
\end{figure}
More explicitly, Fig. \ref{fig:radii} shows, in parameterized form, the evolution of the radial function $z(y)=1/(\Omega_{+} \tilde{W}(y))$ for different values of $\alpha$. The existence of a minimum radius is evident for $|\alpha| \gtrsim 0.8$. A numerical analysis puts forward a linear relation between the wormhole throat radius ($z_{th} = z_{min}$) and the value of $\alpha$ (see Fig. \ref{fig:rminalpha}). With an appropriate constant rescaling of units in the line element, this linear relation can be brought into the more suggestive expression $r_{th}\approx |\alpha|r_\epsilon=2M$, which is valid for all $|\alpha| \gtrsim 0.8$. The coincidence of this value with the location of the event horizon in Schwarzschild black holes of mass $M$ is certainly remarkable and could not have been anticipated before the numerical analysis.

For lighter configurations, we can use an approximated linear form for $l_2$ as a function of $\alpha$, in order to determine the critical value, $\alpha_c$, for which the change of wormhole/non-wormhole regime occurs. Thus, for $|\alpha| \leq 1$, the fitting of the numerical results, see Fig. \ref{fig:l2}, gives $l_ 2 = -0.330976 + 1.00735\, \alpha$, which leads to the critical value $\alpha_c \approx -0.82$. A rough estimate of the corresponding critical Schwarzschild mass gives the result $M_c= -\frac{1}{2} r_\epsilon \alpha_c  \approx 0.41 r_\epsilon \approx 0.013$.

\subsection{Scalar sector}

\begin{figure}[h]
\centering
\includegraphics[width=0.45\textwidth]{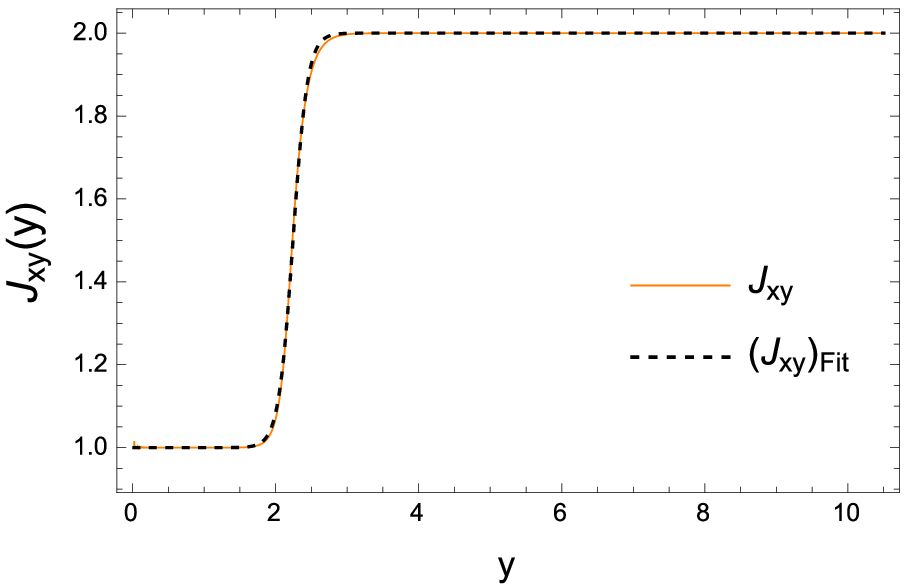}
\caption{The function $J_{xy}$ (solid) and its approximation $(J_{xy})_{Fit}$ (dashed) in Eq.(\ref{eq:Jxy}), for the case $\alpha=-10$.  \label{fig:3}}
\vspace{4mm}
\includegraphics[width=0.45\textwidth]{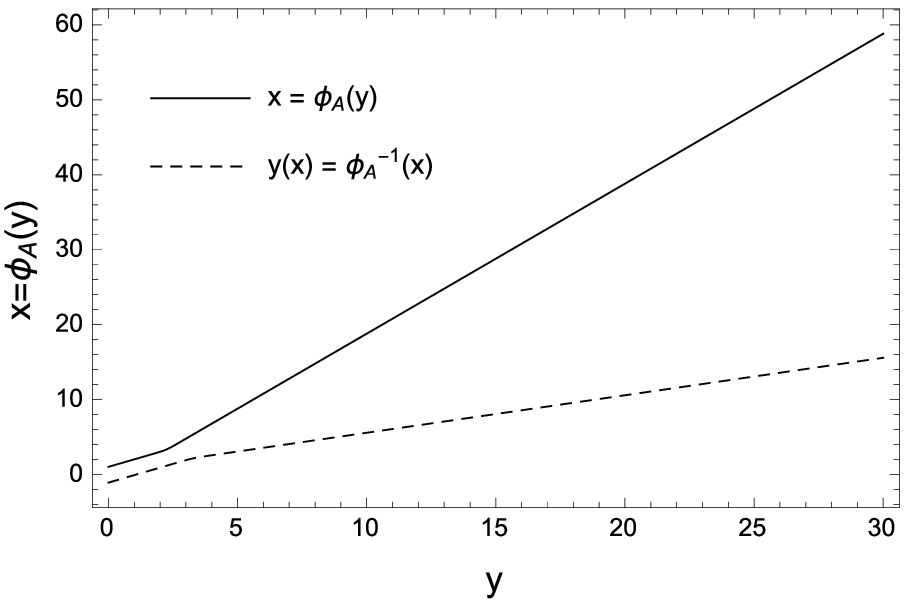}
\caption{Scalar field $\phi_A(y)$ from integration of the approximated function $(J_{xy})_{Fit}$ in Eq.(\ref{eq:phiAy}) (solid) and its inverse (dashed), for the case $\alpha=-10$ and with the integration constant $\phi_0$ set to $0$ for simplicity.}\label{fig:4}
\end{figure}

It is time now to study the behavior of the scalar field. Let us recall that we have identified the radial coordinate with the scalar field, $x \leftrightarrow \phi(x)$. Therefore, from the change of variables \eqref{eq:dydx}, suitably expressed as $dx = J_{xy}\, dy=|1 - X_\epsilon|\, dy = (1 -  X_\epsilon)\,dy$,
we can write $\phi(y)=\int (1 -  X_\epsilon) dy$. This way, an analytical function fitting the numerical solution for $J_{xy}$ can be found as

\begin{equation}
(J_{xy})_{Fit} = 1 + \tfrac{1}{2} \left(1 + \tanh[\alpha\, m_J  (y - y_c)]\right) \ . \label{eq:Jxy}
\end{equation}
In Fig.~\ref{fig:3} we observe that $(J_{xy})_{Fit}$ fits well the numerically integrated function $J_{xy}$ with the parameters depicted in Table \ref{table:III}.
This allows us to integrate the approximating function $(J_{xy})_{Fit}$ so as to obtain the approximated field profile $\phi_A(y)$ as

\begin{equation} \label{eq:phiAy}
\phi_A(y)= \phi_0 + \tfrac{3}{2} y + \frac{1}{\alpha\, m_J} \log [ \cosh(\tfrac12 \alpha\, m_J (y - y_c))] \ ,
\end{equation}
shown in Fig. \ref{fig:4}, along with its inverse $y(x)=\phi_A^{-1}(x)$.

\begin{table}[h]
\begin{center}
\begin{tabular}{c | c c c c}
$\alpha$ &-50 & -10& -1& -0.8 \\[1mm]
\hline
$m_J$ & -0.9593 & -0.9844 &-2.6970&-3.3552\\
\end{tabular}
\caption{Fitting values of the parameter $m_J$ in the approximation (\ref{eq:phiAy}) for different values of $\alpha$.}
\label{table:III}
\end{center}
\end{table}

We will see in the next section that, despite its innocent appearance, the presence of the scalar field gives rise to a remarkable rich and interesting structure.

\section{Solitonic behavior} \label{sec:VI}

When Eq.\eqref{eq:scalarfield} was presented, it was mentioned that the free scalar field is completely determined by the geometry. However, the geometry is also specified by the scalar field and, therefore, the final solution is the result of a nonlinear interaction of the scalar field with itself through the metric it generates. As a consequence, in the discussion of the energy associated to this type of solutions one should take into account not only the energy of the scalar field but also the energy stored in the gravitational field generated by the scalar field. In this sense, the scalar field energy density can be read from the $T_{00}$ component of its stress-energy tensor in Eq.(\ref{eq:tmunus}), which gives  $T_{00}=-\frac{1}{2}g^{xx} \phi_x^2 g_{00}=-\frac{1}{2}g^{xx} g_{00}$, in the representation $\phi_x=1$. Then, using the relations

\begin{eqnarray}\label{rels}
X_\epsilon&=&\frac{\epsilon\kappa^2}{2} C^2 W^4 e^{-\nu} \hspace{0.1cm};\hspace{0.1cm} \sqrt{-g}=C \sin\theta g_{xx} \nonumber \\
&\Rightarrow& \;g_{xx}=\frac{1}{C^2 W^4 e^{-\nu}} =\frac{\epsilon\kappa^2}{2 X_\epsilon} \ ,
\end{eqnarray}
the energy density of the scalar field becomes

\begin{equation} \label{eq:T00scalar}
T_{00}= \frac{-e^{\tilde{\nu}}}{\epsilon\kappa^2} \frac{\left(1- \sqrt{\Omega_+ \Omega_-}\right)}{\Omega_+}
= \frac{-e^{\tilde{\nu}}}{\epsilon\kappa^2} \frac{\left(1- |\lambda+X_\epsilon|\right)}{(\lambda^2- X_\epsilon^2)^{1/2}} \ .
\end{equation}
Using the numerical solution obtained in Eq.(\ref{eq:phiAy}), it turns out that the scalar field energy density per unit time \eqref{eq:T00scalar} is divergent as it approaches the center (corresponding to $y \rightarrow \infty$ and $X_\epsilon\approx -\lambda+\frac{1}{2} |\theta|^{-2/3}$).

An alternative measure of the total energy density, which has been used previously in the context of Born-Infeld gravity coupled to electromagnetic fields with interesting results \cite{or13}, is the spatial part of the integrand of the action functional calculated on the solutions. This quantity can be seen as associated to the energy of the scalar field plus a gravitational binding energy. In the case of a scalar field coupled to GR, this quantity is proportional to the scalar field potential and, therefore, vanishes for free fields. In the Born-Infeld theory, however, the total action of the theory evaluated on the solutions takes the form

\begin{equation}\label{SBIS}
S_{BI+\phi} = \frac{1}{\kappa^2 \epsilon}\int d^4x \left[\sqrt{-q} - \lambda \sqrt{-g}\right] - \frac{1}{2} \int d^4x \sqrt{-g} X \ ,
\end{equation}
(recall that $X\equiv g^{\mu\nu} \partial_\mu\phi\partial_\nu\phi$) and can be rewritten in terms of the deformation matrix ${\Omega^\mu}_{\nu}$ and the relation (\ref{rels}) as

\begin{eqnarray} \label{eq:energydef}
S_{BI+\phi} &=& \frac{1}{\kappa^2 \epsilon}\int d^4x \sqrt{-g} \left[ (|\Omega|^{1/2}-\lambda) - X_\epsilon\right] \nonumber \\
&=& \int d^4x \frac{C \sin\theta}{2 X_\epsilon} \left[ |\Omega|^{1/2}- (\lambda + X_\epsilon)\right] \\
  &=&{2\pi C} \int dt \int dy \frac{ |\lambda^2- X_\epsilon^2|}{X_\epsilon} \left[ (\lambda^2- X_\epsilon^2)^{1/2} - 1 \right] \ , \nonumber
\end{eqnarray}
where we have used the coordinate change (\ref{eq:dydx}). As we are considering free fields, the energy density defined by the Lagrangian density becomes

\begin{equation} \label{eq:energysoliton}
\varepsilon_{BI+\phi}(y) \propto |{X_\epsilon}|^{-1}  |\lambda^2- X_\epsilon^2|\left[ (\lambda^2- X_\epsilon^2)^{1/2} - 1\right] \ ,
\end{equation}
whose behavior can be studied using the analytical and numerical results of previous sections. %approximation above, Eq.(\ref{eq:phiAy}).

\begin{figure}[h]
\centering
\includegraphics[width=0.42\textwidth]{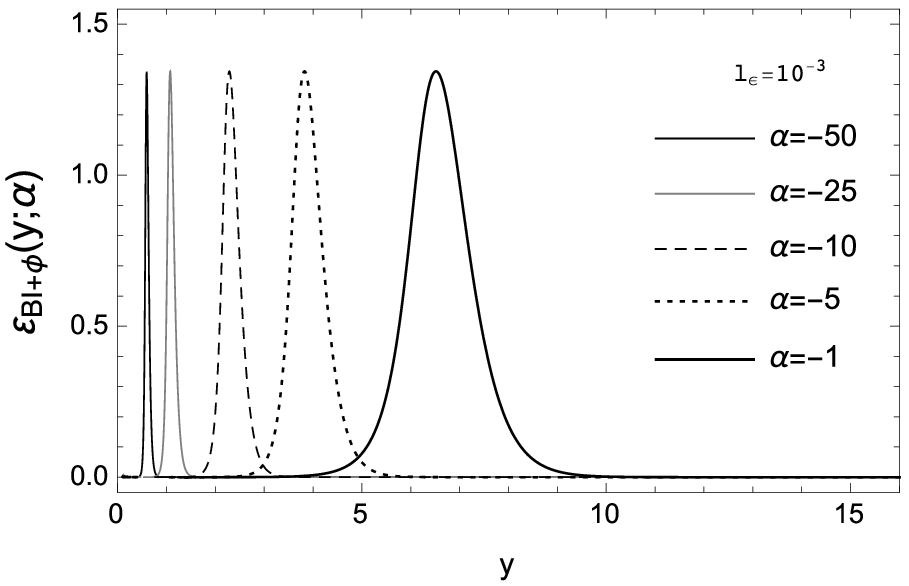}
\vspace{-2mm}\caption{The total energy density for the combined system gravity + matter, $\varepsilon_{BI+\phi}(y)$, in Eq.(\ref{eq:energysoliton}), with Born-Infeld parameter $l_{\epsilon}=10^{-3}$ and setting $\lambda=1$, for several values of the integration constant $\alpha$.}\label{fig:7}
\vspace{5mm}
\includegraphics[width=0.42\textwidth]{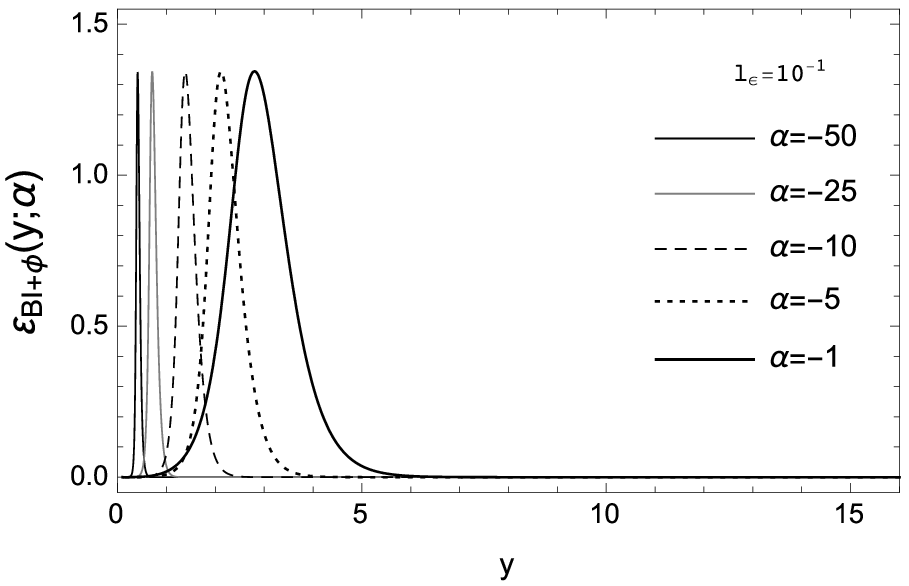}
\vspace{-2mm}\caption{Same notation as in Fig.~\ref{fig:7}, now with $l_{\epsilon}=10^{-1}$. As compared to the previous plot, now the solutions are localized farther from the center ($y \rightarrow \infty$).} \label{fig:7b}
\end{figure}

In this sense, as depicted in Fig.~\ref{fig:7}, for any value of the integration constant $\alpha$ in Eq.(\ref{eq:nufit}) the total energy density presents a characteristic profile made up of a \emph{lump} of energy localized in a finite region of space, while it vanishes both in the asymptotic limit, $\theta \rightarrow 0$, and in the central region, $\theta \rightarrow \infty$. As already discussed in the introduction, the localization of the energy density (scalar plus gravitational in this case) in a finite region of space is a characteristic feature of solitonic configurations. Note that the variation of the integration constant $\alpha$ makes the localization of the lump of energy density to be shifted and its width to be changed, though both the localized nature and the amplitude of the lump remain unchanged.

Comparing the calculations made with $l_{\epsilon}=10^{-3}$ and $l_{\epsilon}=10^{-1}$ in Figs.~\ref{fig:7} and \ref{fig:7b}, respectively, we see that the effect of increasing the relative strength of the gravitational Born-Infeld corrections, encoded in $l_{\epsilon}$, is to shift the location of the maximum of the energy density farther from the center of the solution (located at $y \rightarrow \infty$) and to increase its width, while its height remains unchanged. None of these solitonic features is found for the GR counterpart (corresponding to $l_{\epsilon} \rightarrow 0$).

\begin{figure}[h!]\centering
\includegraphics[width=0.45\textwidth]{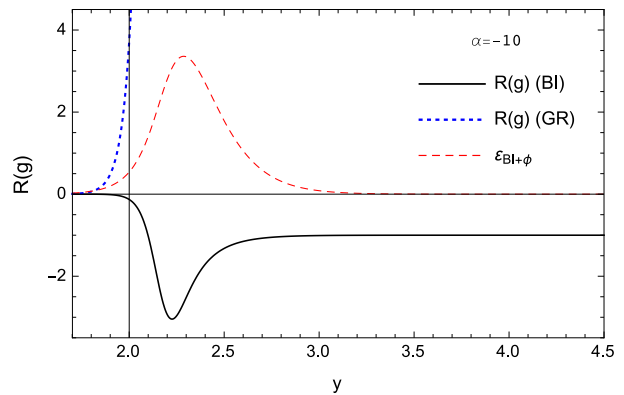}
\caption{Normalized Ricci scalar of the spacetime metric $g_{\mu\nu}$ in both Born-Infeld gravity (solid black) and GR (dotted blue) cases. The (total) energy density distribution of the gravitational and scalar field sectors (dashed red) defines the region where the curvature evolves from a negative constant value in the internal region ($y \rightarrow\infty$) to zero in the asymptotically flat region ($y \rightarrow 0$).}
\label{fig:Rsy}
\end{figure}

\begin{figure}[!h]\centering
\includegraphics[width=0.45\textwidth]{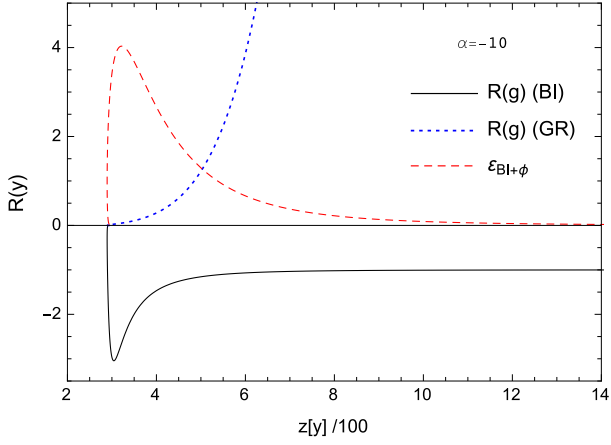}
\caption{Parametric plot of the normalized Ricci scalar of the physical metric $g_{\mu\nu}$ against the radial function $z$ comparing Born-Infeld gravity (solid black) and GR (dotted blue) cases. The energy density distribution (dashed red) defines the region of curvature change from zero (flat) at the minimum of the radial function, to a negative constant value in the internal region. The minimum of the radial function occurs at $z\sim 300$ for $\alpha=-10$, as can be seen from Fig. \ref{fig:radii}.}
\label{fig:Rsz}
\end{figure}

\begin{figure}[!h]\centering
\includegraphics[width=0.45\textwidth]{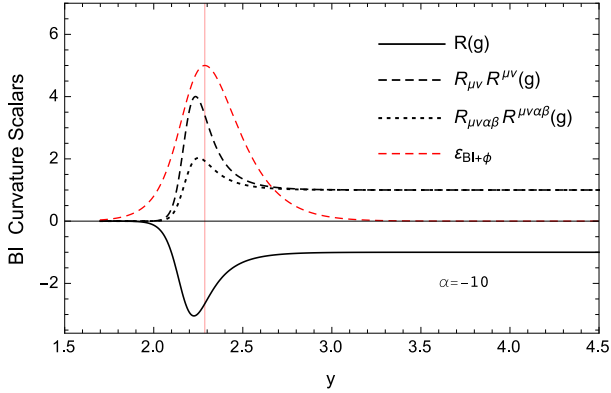}
\caption{Curvature scalars for the Born-Infeld gravity model: Ricci scalar (solid), squared Ricci tensor (dashed) and squared Riemann tensor (dotted) showing the change in the curvature scalars in the region of energy density localization.}
\label{fig:scalarsy}
\end{figure}

\begin{figure}[!h]\centering
\includegraphics[width=0.45\textwidth]{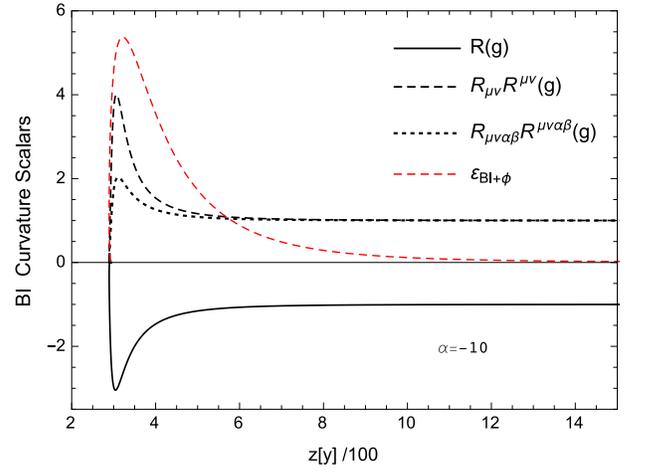}
\caption{Parametric plot of the normalized curvature scalars for the Born-Infeld gravity model: Ricci scalar (solid), squared Ricci tensor (dashed) and squared Riemann tensor (dotted). The energy density (dashed red) localization region signals the curvature change from zero (flat) at the minimum of the radial function, to constant values in the internal region.}
\label{fig:scalarsz}
\end{figure}

\begin{figure}[!h]\centering
\includegraphics[width=0.42\textwidth]{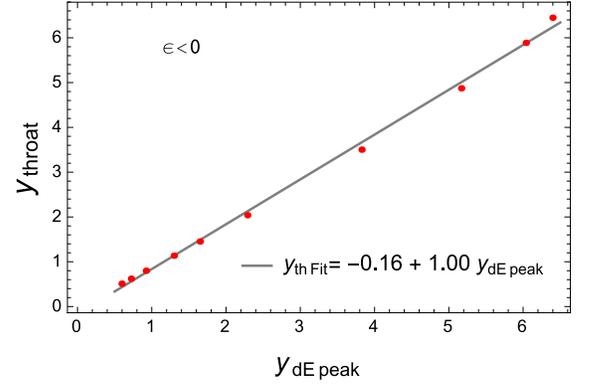}
\vspace{-2mm}\caption{Linear relation between the location of the wormhole throat and the location of the peak of the total energy density for $\epsilon<0$. Within the numerical accuracy, they are coincident.} \label{fig:zWHzpeak}
\end{figure}

To get further into this interpretation we can take a glance at the behavior of the curvature scalars. As depicted in Figs. \ref{fig:Rsy}, \ref{fig:Rsz}, \ref{fig:scalarsy} and \ref{fig:scalarsz}, there is indeed a correlation between the lump of energy and such scalars, as the latter take their maximum value approximately at the center of the lump. This result strongly deviates from the GR case, where nothing in the curvature scalars tells us about the existence of scalar solitonic structures. A glance at Fig. \ref{fig:zWHzpeak} shows that the peak of the lump is localized at the wormhole throat. It must be noted that also in the cases without wormhole the lump localizes on a thick shell away from the center.

\section{Case with $\epsilon>0$} \label{sec:VII-B}

\begin{figure}[h]
\begin{tabular}{c c}
\includegraphics[width=4.0cm, height=3.5cm]{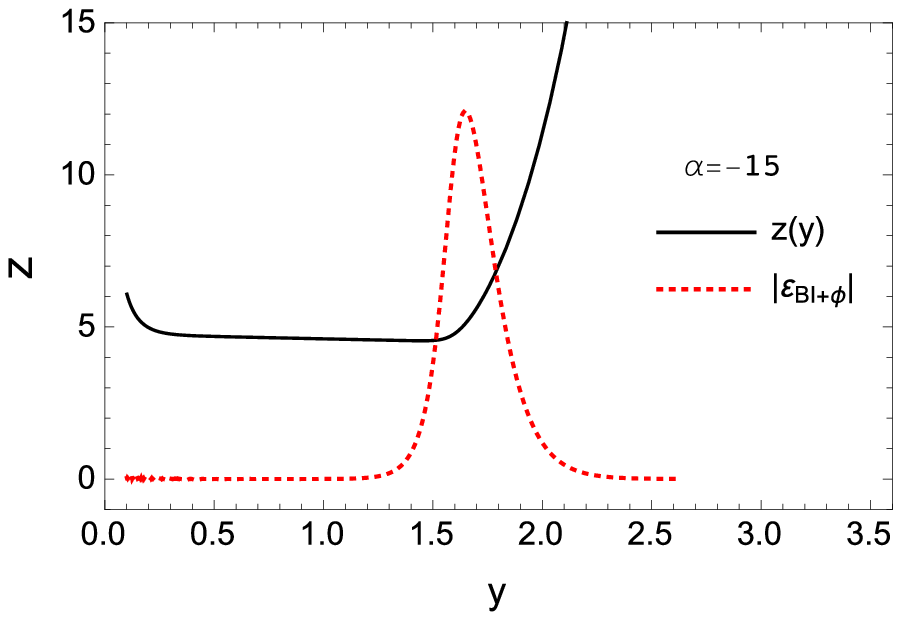}
& \includegraphics[width=4.0cm, height=3.5cm]{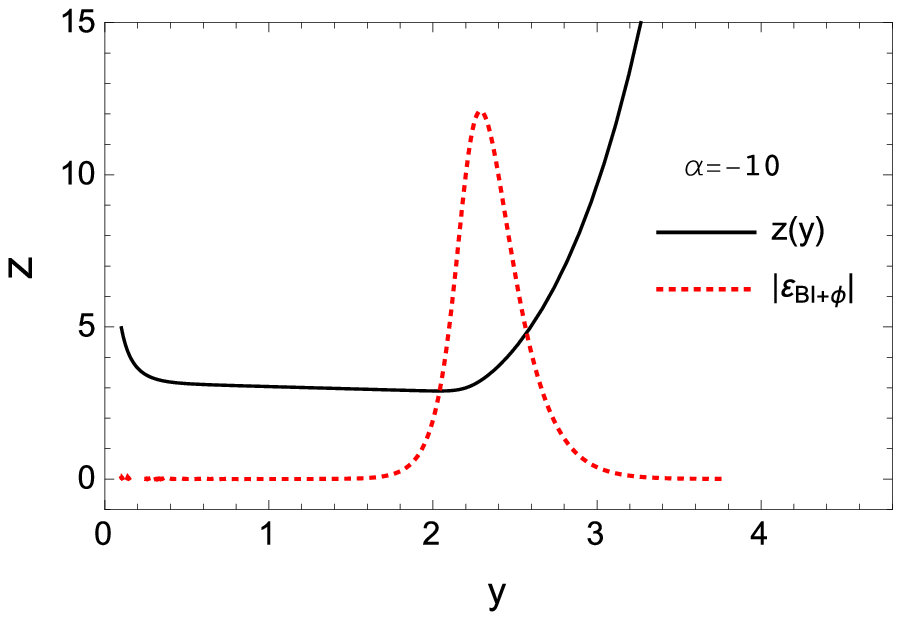} \\
\includegraphics[width=4.0cm, height=3.5cm]{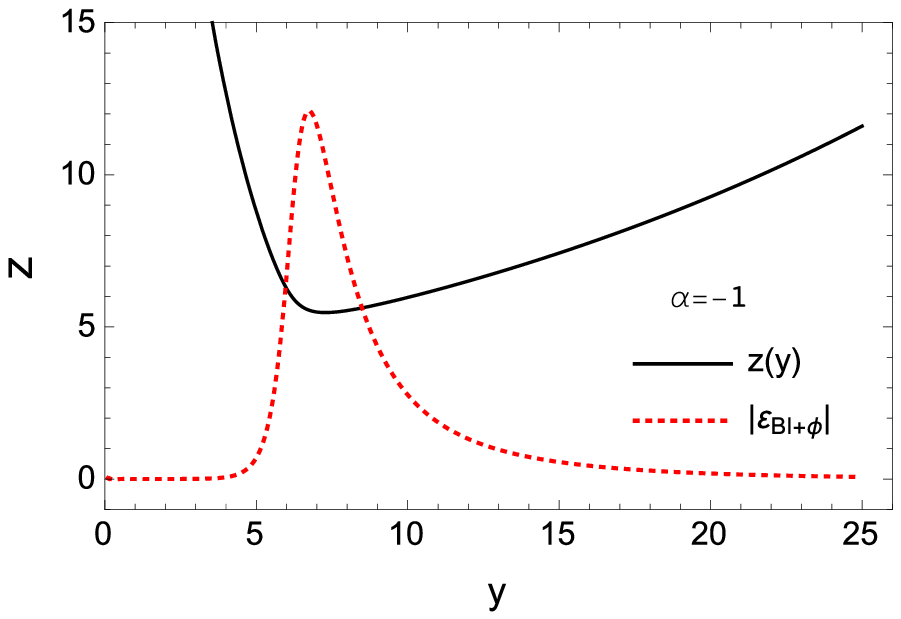}
& \includegraphics[width=4.0cm, height=3.5cm]{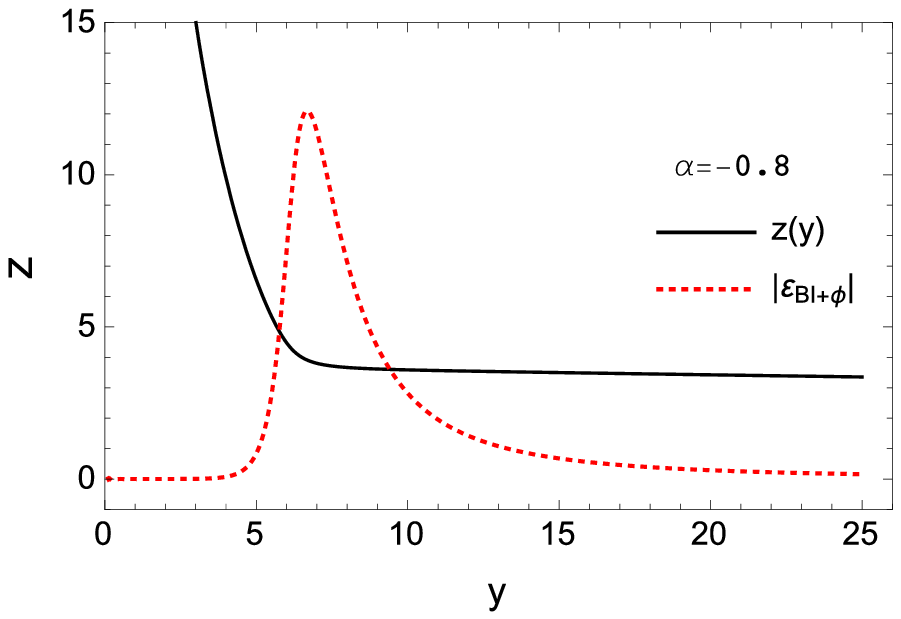}
\end{tabular}
\caption{Radial function $z=1/(r_\epsilon \Omega_{+} \tilde{W}(y))$, and the total energy density $\vert \varepsilon_{BI+\phi} \vert$, as a function of the coordinate $y$ for different values of the parameter $\alpha$. The limits $\vert \theta \vert \rightarrow 0$ and $\vert \theta \vert \rightarrow \infty$ determine the asymptotic (GR-like) and central regions, respectively. For $\alpha \lesssim -0.8$ a minimum in the radial function is found, which signals the presence of a wormhole structure.
\label{fig:WHen}}
\end{figure}

\begin{figure}[h]\centering
\includegraphics[width=0.45\textwidth]{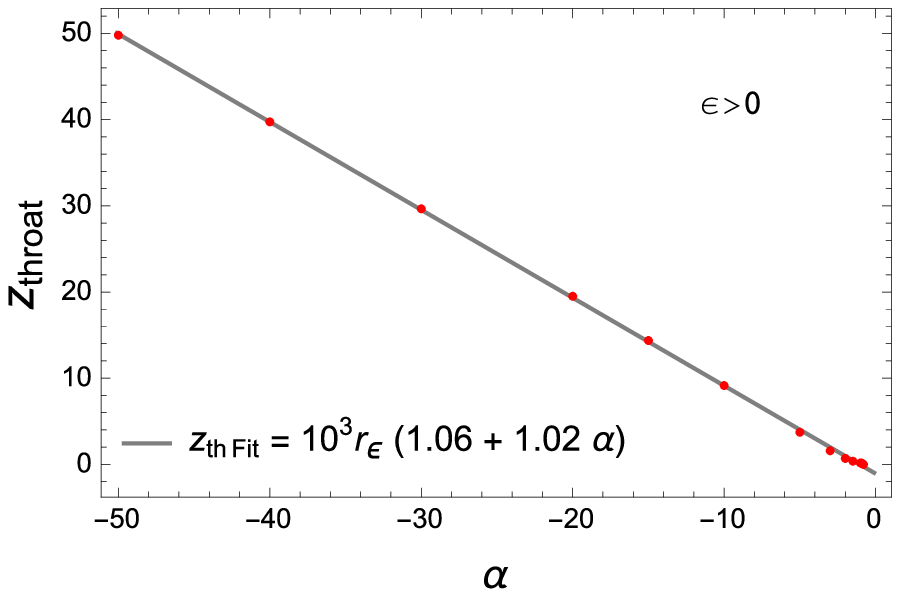}
\caption{Linear relation between the location of the wormhole throat and the parameter $\alpha=-2M/r_\epsilon$ in the case $\epsilon>0$. }
\label{fig:rminalphaEpsP}
\end{figure}

In sections \ref{sec:V} and \ref{sec:VI}  above, we have solved the field equations (\ref{fulleq1}), (\ref{fulleq2}) and (\ref{fulleq3}) under the assumption $\epsilon=-2l_{\epsilon}^2<0$ and discussed the physical properties of the solutions. For completeness, in this section we shall study the branch $\epsilon>0$ and compare the results of both cases. Let us first consider the two regions of interest. In this sense, in the asymptotic limit, $ \vert \theta \vert \rightarrow 0$, we find that these equations boil down to

\begin{eqnarray}
X_\epsilon &\simeq& |\theta| \approx 0 \ , \\
\tilde{\nu}_{yy} &\simeq & \frac{\kappa_0^2}{2} |\theta| \approx 0 \ , \\
\tilde{W}_{yy}-\tilde{\nu}_y\tilde{W}_y &\simeq& \frac{\kappa_0^2}{2} (1 - |\theta|) \tilde{W} \approx   \frac{\kappa_0^2}{2} \tilde{W} \ , \\
\tilde{W}\tilde{W}_{yy}-\tilde{W^2_y}+\frac{e^{\tilde{\nu}}}{C_0^2} &\simeq& \frac{\kappa_0^2}{4} |\theta| \, \tilde{W}^2 \approx 0 \ .
\end{eqnarray}
which recover the GR behavior there, like in the $\epsilon <0$ case, see Eqs.(\ref{Wy1}), (\ref{Wy2}) and (\ref{Wy3}). In the central region, $\vert \theta \vert \rightarrow \infty$, we find instead

\begin{eqnarray}
X_\epsilon&\simeq &  (1 - \tfrac{1}{2} |\theta|^{-2/3}) \approx 1 \ , \\
\tilde{\nu}_{yy}&\simeq& \kappa_0^2 \,|\theta|^{-2/3} \approx 0 \ , \\
\tilde{W}_{yy}-\tilde{\nu}_y\tilde{W}_y & \simeq &   \frac{\kappa_0^2}{4} \tilde{W} |\theta|^{-2/3} \approx 0 \ , \label{eq:Wsp} \\
\tilde{W}\tilde{W}_{yy}-\tilde{W^2_y}+\frac{e^{\tilde{\nu}}}{C_0^2} &\simeq & - \frac{1}{2} \kappa_0^2 |\theta|^{-2/3} \tilde{W}^2
 \approx 0 \ .
\end{eqnarray}
It is immediately seen that the equation (\ref{eq:Wsp}) for $\tilde{W}$ gets simplified as compared to the case $\epsilon<0$, see Eq.(\ref{fulleq2}), which in turn modifies the inner behavior of the solutions. Indeed, now it is possible to analytically integrate the equations at the center of the solutions as

\begin{eqnarray}
\tilde{\nu}_{Center}(y)&=& L_1+ L_2\, y \ ,\\
\tilde{W}_{Center}(y)&=&\frac{e^{L_1}}{L_2 D_1 C_0^2} e^{L_2\,y} -\frac{D_1}{L_2} \label{eq:Wep2} \ ,
\end{eqnarray}
where $L_1$ and $L_2$ are integration constants to be determined by means of the numerical computation, and we have defined $D_1=\tilde W'(y) - L_2 \, \tilde W(y)$. Using the asymptotic limits above we have implemented a similar numerical strategy as in the $\epsilon<0$ case. As the most relevant results, we find again evidence of existence of wormhole structures, again for $\alpha \lesssim -0.8$, see Fig. \ref{fig:WHen}, for which the radial function $z(y)$ bounces off to $z \rightarrow \infty$ after reaching a minimum (the throat). As in the case with $\epsilon<0$, for $\epsilon>0$ the location of the throat is also determined by the mass of the object (see Fig. \ref{fig:rminalphaEpsP}). Such structures disappear for objects with lighter masses ($\alpha$ closer to zero), which is a similar result as in the $\epsilon <0$ case.

Attending to the relation $z^2(y)\approx\left(\tilde{W}^2e^{\tilde{\nu}}\right)^{-\frac{1}{3}}$ that arises in the $|\theta|\to \infty$ limit, the existence of wormhole/non-wormhole structures is justified by a change of sign in the parameter $L_2$ around $\alpha \simeq -0.8$. In fact, for $\alpha <-0.8$ we observe $L_2<0$ as $y\to\infty$, which leads to a minimal $2$-sphere or radius $\tilde{z}_{min}=\lim_{y\to \infty}1/\tilde{W}=-L_2/D_1$ for the auxiliary metric. Consequently, $z^2(y)\propto e^{-L_2y/3}$ grows with increasing $y$, corresponding to a wormhole. On the other hand, when $L_2>0$ one finds that $\tilde{z}\propto e^{-L_2 y}$ goes to zero as $y$ grows. In this case, $L_2>0$ implies that $z^2(y)\propto e^{-L_2 y}\to 0$ as $y\to \infty$.

\begin{figure}[h]
\centering
\includegraphics[width=0.45\textwidth]{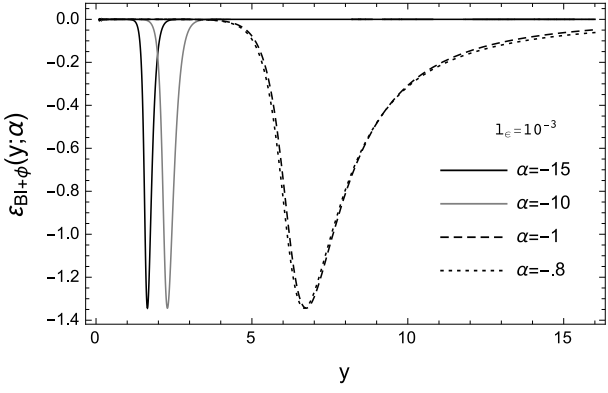}
\caption{The total energy density for the combined system gravity + matter, $\varepsilon_{BI+\phi}(y)$ in Eq.(\ref{eq:energysoliton}), with Born-Infeld parameter $l_{\epsilon}=10^{-3}$, for several values of the parameter $\alpha$. When a wormhole exists, the minimum of the curve coincides with the location of the throat. \label{fig:wall}}
\end{figure}

\begin{figure}[h]
\centering
\includegraphics[width=0.45\textwidth]{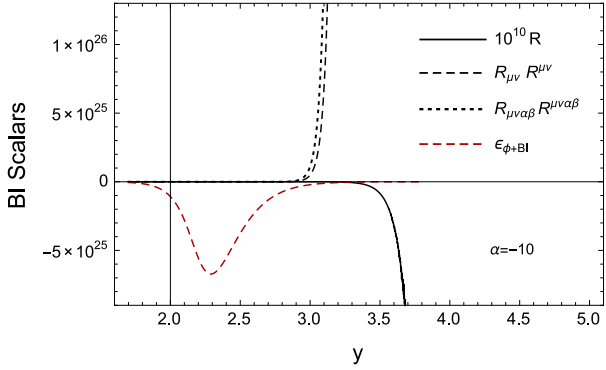}
\caption{Behavior of the curvature scalars for $\epsilon>0$, compared to the total energy density. \label{fig:csen}}
\end{figure}

In the present case, computation of the total energy density made up of the gravitational and scalar field contributions, Eq.(\ref{eq:energydef}), for wormhole and non-wormhole configurations, yields again a localized profile at a finite distance, but now it represents a \emph{well} instead of a lump, as such an energy density is negative [see Fig. \ref{fig:wall} and the definition of total energy density in Eq.(\ref{eq:energydef})]. Unlike in that case, the location of the peak is not related to extrema of the curvature scalars, see Fig. \ref{fig:csen}, though it coincides with the wormhole throat when it exists.

Curvature scalars can be indeed computed using the generic form (\ref{eq:ds2thetainf0}) obtained for the line element when $|\theta|\to\infty$ but taking into account that for  $\epsilon>0$ we have  $\lim_{|\theta|\to \infty}dx^2\approx dy^2/(4|\theta|^{4/3})$. As a result, the line element takes the form
\begin{equation}\label{eq:ds2epplusz}
\frac{ds^2}{r_\epsilon^2}=-\frac{1}{z^4}dt^2+\frac{\sigma^2L_2^2}{4C_0^2D_1^2}\frac{dz^2}{z^{10}}+z^2d\Omega^2 \ ,
\end{equation}
with curvature scalars given by
\begin{eqnarray}
r_\epsilon^4{R^\alpha}_{\beta\mu\nu}{R_\alpha}^{\beta\mu\nu} &\approx & \frac{4}{z^4} -\left( \frac{4C_0^2D_1^2}{\sigma^2L_2^2}\right)8z^6 \\
&&+\left(\frac{4C_0^2D_1^2}{\sigma^2L_2^2}\right)^2 300 z^{16}\nonumber  \\
r_\epsilon^2R&\approx&\frac{2}{z^2} -\left( \frac{4C_0^2D_1^2}{\sigma^2L_2^2}\right)6 z^8 \ ,  \label{eq:scalarsepsp}
\end{eqnarray}
which are divergent both in the $z\to0$ and $z\to \infty$ limits.

\section{Geodesics} \label{sec:VII}

Given that in the cases studied above the geometry far from where the scalar field is localized is asymptotically flat and recovers that of GR, the discussion of geodesics is only relevant in the interior region ($y\to \infty$). For this reason, in this section we consider the analytical approximations obtained so far to discuss their properties in both GR and the Born-Infeld gravity cases. This will also allow us to discuss the regularity of these spacetimes attending to the criterion of geodesic completeness, namely, whether any geodesic curve can be extended to arbitrarily large values of their affine parameter, and which plays a fundamental role in the singularity theorems  \cite{Theorems,SG15}. This captures the intuitive notion that in a physically consistent spacetime neither information nor physical observers (idealized as null and time-like geodesics, respectively) should be allowed to suddenly disappear or emerge from nowhere.

We point out that, in application of Einstein's equivalence principle, which dictates that test particles follow geodesics of the spacetime metric $g_{\mu\nu}$, in the action (\ref{SBIS}) describing our theory the matter does not couple directly to the connection (this is obvious for scalar fields). For this reason, we will focus on the  geodesics of the metric $g_{\mu\nu}$.
In general, should an explicit dependence on the connection in the matter sector be allowed, then one would be led to consider the geodesics associated to the independent connection as physically meaningful (see \cite{LecOlmo} for an extended discussion on geodesics in metric-affine spaces).

The above discussion implies that a geodesic curve $\gamma^{\mu}=x^{\mu}(u)$, where $u$ is the affine parameter,
extremizes the functional \cite{Wald,Chandra}

\begin{equation}
\mathcal{S}=\frac{1}{2} \int du \sqrt{g_{\mu\nu} \frac{dx^{\mu}}{du}\frac{dx^{\nu}}{du}} \ .
\end{equation}
This means that $x^{\mu}(u)$ must satisfy the equation

\begin{equation}
\frac{d^2 x^\mu}{d u^2}+\Gamma^\mu_{\alpha\beta}(g) \frac{d x^\alpha}{d u}\frac{d x^\beta}{d u}=0 \ ,
\end{equation}
where $\Gamma^\mu_{\alpha\beta}(g)$ are the Christoffel symbols associated to the spacetime metric $g_{\mu\nu}$. In the case of a spherically symmetric metric of the form $ds^2=-C(x) dt^2+D(x)^{-1} dx^2+r^2(x)d\Omega^2$ the geodesic equation takes the form \cite{LecOlmo}

\begin{equation} \label{eq:geoeqo}
\frac{C(x)}{D(x)} \left(\frac{dx}{du}\right)^2=  E^2 -C(x) \left(\frac{L^2}{r^2(x)} -k \right),
\end{equation}
where the parameter $k=1,0,-1$ corresponds to space-like, null and time-like geodesics, respectively. For time-like geodesics, the conserved quantities $E=\sqrt{DC} dt/du$ and $L=r^2(x)d\varphi/du$  (due to staticity and spherical symmetry) have the interpretation of the total energy per unit mass and angular momentum per unit mass, respectively, around an axis normal to a plane (which can be chosen to be $\theta=\pi/2$ without loss of generality). For null geodesics this interpretation cannot be sustained, but the quotient $L/E$ can be identified instead as an apparent impact parameter from asymptotic infinity \cite{Chandra}.

Using the line elements (\ref{eq:ds2center}) and (\ref{eq:ds2epplusz}), the above expression (\ref{eq:geoeqo}) yields
\begin{equation} \label{eq:geoeqo3}
\frac{a_\epsilon^2}{z^{4+2n_\epsilon}} \left(\frac{dz}{du}\right)^2=  E^2 -\frac{1}{z^4} \left(\frac{L^2}{z^2} -k \right) ,
\end{equation}
where $a_\epsilon$ and $n_\epsilon$ are defined as
\begin{equation}
(a_\epsilon,n_\epsilon)=\left\{
\begin{array}{ll}
\left(\tfrac{1}{12}C_0(2l_2+l_\kappa),1\right) & \text{ if } \epsilon<0 \\
\left(C_0D_1,5\right) & \text{ if } \epsilon>0 \ , L_2>0 \\
\left(\tfrac{1}{3}C_0D_1,5\right) & \text{ if } \epsilon>0 \ , L_2<0
\end{array}\right. \\
\end{equation}
To proceed with the discussion of geodesics, we must now distinguish between wormhole and non-wormhole configurations.

\subsection{Non-wormhole case}

In the non-wormhole case, %($\vert \alpha \vert$ small as compared to $2\kappa/\sqrt{3}$, see Sec.\ref{sec:secSS})
in which $z\to 0$ as $y\to\infty$, it is easy to see that for geodesics with angular momentum $L\neq 0$ (and any value of $k$), the right-hand side of \eqref{eq:geoeqo3} must necessarily vanish  at some finite and small radius as $z$ approaches zero. This means that such geodesics attain a minimum and then bounce to increasing values of the radial coordinate. This way they, evidently, never reach the center.

When $L=0$, time-like geodesics also bounce. On the other hand, for radial null geodesics ($L=0=k$), Eq.\eqref{eq:geoeqo3} can be easily integrated leading to

\begin{equation}\label{eq:null_rad_BI}
\frac{a_\epsilon}{1+n_\epsilon}\left(\frac{1}{z_0^{1+n_\epsilon}}-\frac{1}{z^{1+n_\epsilon}}\right)=\pm E(u-u_0) \ ,
\end{equation}
where $u_0$ is an integration constant, and the $\pm$ sign represents outgoing/ingoing geodesics, respectively. It is easy to see that ingoing geodesics take an infinite affine time to reach the origin. Similarly, the initial condition at the center for outgoing geodesics must be set at $u\to -\infty$. This result puts forward that these spacetimes are geodesically complete, with the center located beyond the reach of geodesic observers and light rays. The curvature divergences that arise at $z\to 0$ [see Eqs.(\ref{eq:scalars}) and (\ref{eq:scalarsepsp})] are thus inaccessible. This situation is reminiscent of that found in some recently studied Palatini $f(R)$ theories sourced by anisotropic fluids \cite{fRani} or electromagnetic fields \cite{fRem}, with the interesting result that the GR point-like singularity is generically replaced by a finite-size wormhole structure. Though curvature divergences appear at the wormhole throat, this structure lies on the future (or past) boundary of the spacetime and cannot be reached in finite affine time. Note also that for spatial geodesics (with $L=0$), the geodesic equation leads to $u\propto \ln z$ when $\epsilon<0$ and to $u\propto 1/z^4$ for $\epsilon>0$ . This implies that $z\to 0$ is at an infinite affine distance and confirms that also spatial geodesics are complete.

In order to facilitate the comparison with GR, we note that in the limit $y\to\infty$ the geodesic equation takes the form

\begin{equation}\label{eq:geodesicsGR}
\left(\frac{\gamma}{m_+}\left(\frac{\gamma}{z}\right)^{\frac{\alpha}{m_+}-1}\right)^2\left(\frac{dz}{du}\right)^2=
E^2-\left(\frac{\gamma}{z}\right)^{\frac{\alpha}{m_+}}\left(\frac{L^2}{z^2}-k\right) \ .
\end{equation}
Given that to get the correct weak field limit we need $\alpha=-2M/r_\epsilon<0$ and that $m_+>0$, we see that the factor $\left(\frac{\gamma}{z}\right)^{\frac{\alpha}{m_+}}$ on the right-hand side goes to zero as $z\to 0$. Moreover, for sufficiently massive objects one has $|\alpha|/m_+ \approx \gamma/m_+ \gg 1$, which implies that time-like and $L\neq 0$ geodesics behave like radial null geodesics near the origin. For radial null geodesics in the approximation above\footnote{For radial null geodesics in GR, an exact analytical expression with the form $y=y(\lambda)$ is possible.}, equation \eqref{eq:geodesicsGR} can be integrated to get

\begin{equation}
\gamma^{-\frac{2\gamma}{m_+}}\left({z}^{\frac{2\gamma}{m_+}}-z_0^{\frac{2\gamma}{m_+}}\right)=\pm 2E (u-u_0) \ .
\end{equation}
It is clear from this expression that nothing prevents these GR geodesics from reaching the curvature divergence at the origin in a finite affine time. The contrast with the Born-Infeld gravity case is thus remarkable.

\subsection{Wormhole case} \label{sec:WHcase}

In the wormhole case, the function $z(y)$ diverges as $y\to \infty$. This means that, regardless of the values of $k$ and $L$, Eq.(\ref{eq:geoeqo3}) tends to the case of radial null geodesics of Eq.(\ref{eq:null_rad_BI}) but with $z\to \infty$ instead of $z\to 0$. The difference is significant, as now all geodesics can reach the asymptotic infinity in a finite amount of (affine) time. The wormhole, therefore, gives rise to a completely different scenario. It is worth noting that the property of reaching the boundary $z\to \infty$ in a finite affine time is independent of the behavior of curvature invariants. In fact, for $\epsilon<0$ this region has finite invariants, whereas for $\epsilon>0$ they are divergent. The lack of correlation between curvature scalars and geodesic structure seems to be quite generic in theories beyond GR \cite{fRani}.

\subsection{Numerical analysis and effective potentials}

\begin{figure}[h!]\center
\includegraphics[width=0.45\textwidth]{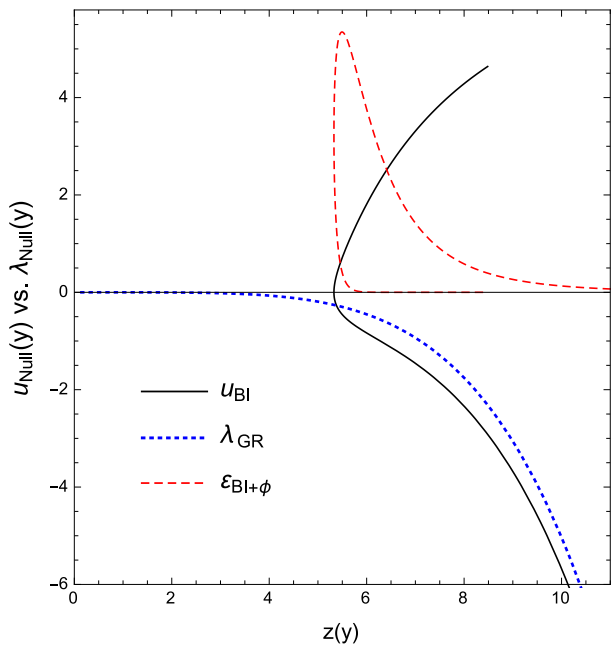}
\caption{Affine parameter for radial null geodesics ($\epsilon<0$, $l_{\epsilon}=10^{-3}\,; \alpha=-1$), superimposed to the total energy density distribution (dashed red). The location of the energy density peak determines also the position of the throat of the Born-Infeld gravity wormhole structure (solid black),  absent in the GR case (dotted blue). The bounce in the trajectory actually represents the crossing of the wormhole throat.
}\label{fig:10}
\end{figure}

In the previous subsections we have studied the asymptotic behavior of geodesics in the internal region according to approximate analytical expressions obtained before. The matching of these asymptotic curves with their GR limit in the external region can be obtained numerically. In Fig. \ref{fig:10}, for instance, the trajectory of a radial null geodesic that goes from the GR region through the wormhole is shown. Non-radial null geodesics in wormhole configurations may go through the wormhole or bounce, depending on their initial energy. In order to quickly visualize the many situations one may encounter, it is useful to plot the effective potentials associated to various configurations.  The effective potential we are referring to appears on the right-hand side of (\ref{eq:geoeqo}) and can be written as
\begin{equation} \label{eq:poteppgen0}
V_{eff}=e^{\tilde \nu} \left( L^2  \tilde W^2 -\frac{k}{\Omega_+} \right) \ .
\end{equation}

\begin{figure}[h!]
\centering
\includegraphics[width=0.41\textwidth]{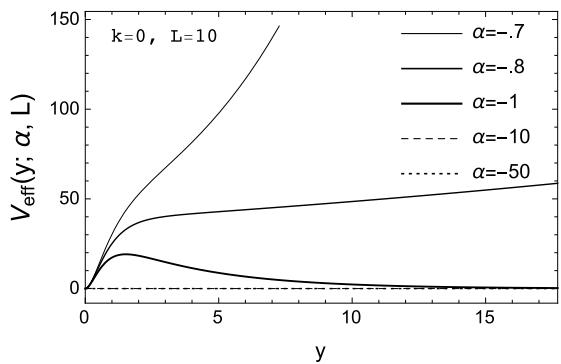}
\caption{Effective potential for non-radial null geodesics, with $L=10$, in the $\epsilon<0$ case. Qualitatively, this shape is the same for the cases $\epsilon>0$ and $\epsilon<0$.
\label{fig:VeffEposL10null}}
\end{figure}

For null geodesics ($k=0$) the above potential simply reads
\begin{equation}
V_{eff}^{rad}
=\left\{\begin{array}{ccl}
0 & &\ L=0 \,\text{(radial)}\\ L^2 {e^{\tilde\nu}} {\tilde W}^2& & \hspace{0.1cm} L\neq0 %\,\text{(non-radial)}
\end{array}\right. \ ,
\end{equation}
which presents similar profiles for all non-radial cases. Moreover, it can be verified that these profiles are qualitatively similar in the $\epsilon>0$ and $\epsilon<0$ branches. To illustrate the general behavior, the case with $L=10$ is shown in Fig. \ref{fig:VeffEposL10null}. There we observe again the transition between the wormhole ($\vert \alpha \vert$ large) and non-wormhole ($\vert \alpha \vert$ small) cases, such that in the former case those geodesics with enough energy to overcome the maximum of the potential barrier will be able to go through the wormhole throat and reach $y \rightarrow \infty$ in finite affine time (this was shown above analytically), while in the non-wormhole case any geodesic will bounce at some finite distance from the center due to the growing barrier as $y\to\infty$, thus corresponding to geodesically complete solutions.

\begin{figure}[t]
\begin{tabular}{c c}
\includegraphics[width=4.2cm, height=3.8cm]{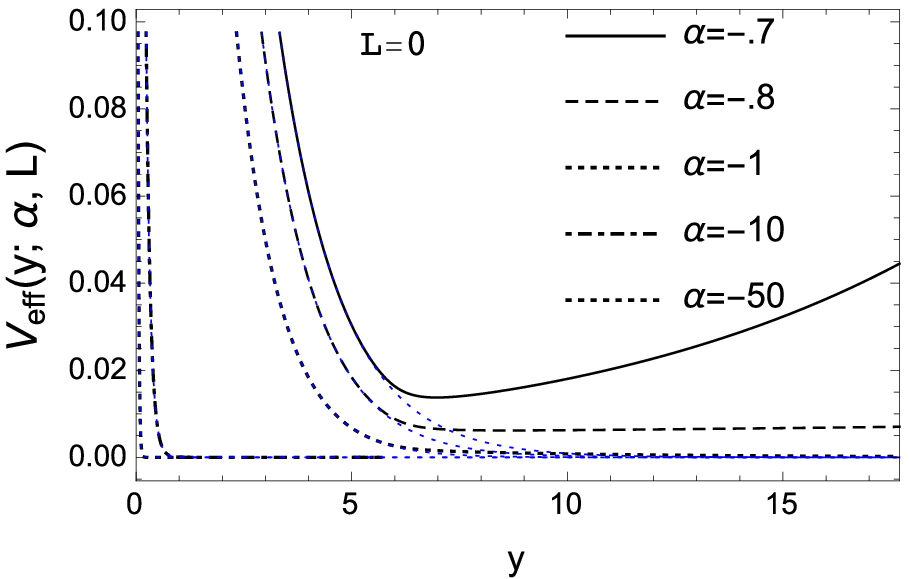}
& \includegraphics[width=4.2cm, height=3.8cm]{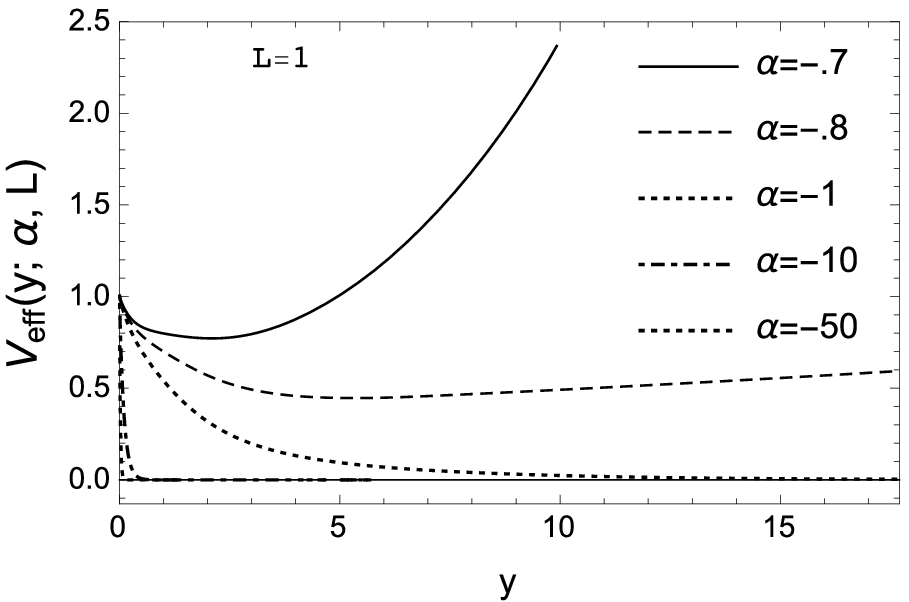} \\
\includegraphics[width=4.2cm, height=3.8cm]{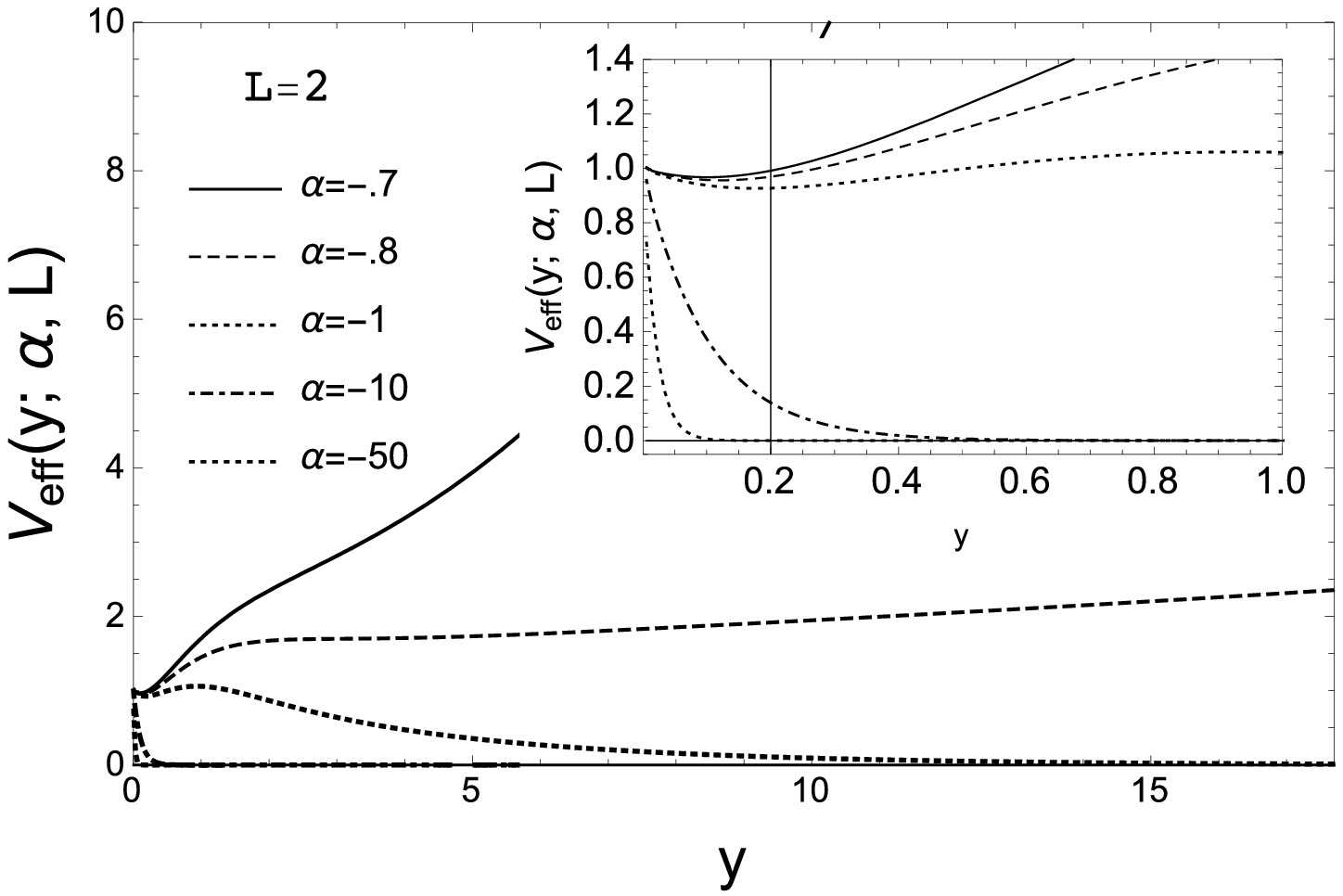}
& \includegraphics[width=4.2cm, height=3.8cm]{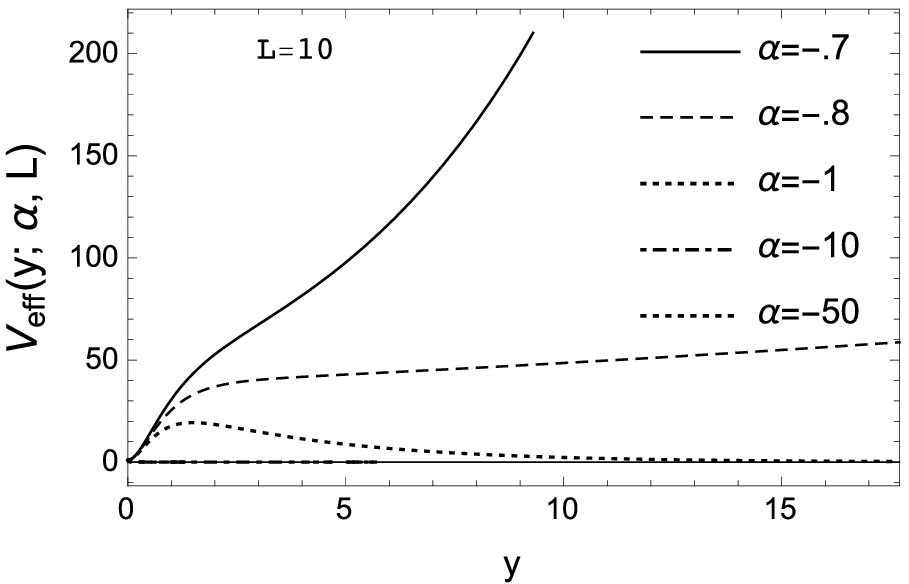}
\end{tabular}
\caption{Effective potential for time-like geodesics ($k=-1$) with different values of $L$ for $\epsilon>0$. Like in the case of null geodesics, these plots are qualitatively similar to those obtained for $\epsilon<0$. In the radial case (left upper panel) we also show the GR behavior (dotted blue lines) for comparison.
\label{fig:vtimeep}}
\end{figure}

For time-like geodesics ($k=-1$) the shape of the potential (\ref{eq:poteppgen0}) is depicted in Fig. \ref{fig:vtimeep} for different values of the angular momentum $L$. The difference between wormhole and non-wormhole configurations is evident from the plots, with the latter exhibiting a growing trend as $y\to\infty$.  For $L\neq 0$, the non-wormhole potentials may lead to minima in which particles could remain stable against radial perturbations. For the wormhole configurations, if the particle has enough energy to surpass the potential barrier near the throat, the interior region offers no resistance at all to its propagation, as $V_{eff}\to 0$ rapidly as $y$ grows. If the particle has not enough energy, then it will bounce before reaching the wormhole and remain in the GR external region.

\section{Conclusions and discussion} \label{sec:VIII}

In this work we have investigated the existence of scalar geonic configurations, namely, self-gravitating solutions supported by a massless, free scalar field. For this purpose we have considered a well motivated extension of GR, dubbed Born-Infeld gravity, characterized by a single length-squared parameter, and which recovers the dynamics and solutions of GR in the low energy-density limit.

After working out the field equations in suitable form for its analysis, we have used a combination of analytical approximations and numerical computations to solve them for the gravitational sector and the scalar field. Two relevant features have emerged out of this analysis. The first one is the existence of two different kinds of structures depending on the value of the constant $\alpha=-2M/r_{\epsilon}$ (which measures the relative size between the scale of the Born-Infeld gravity corrections, $r_{\epsilon}$, and the  Schwarzschild radius associated to the particular solution considered). In this sense, for $\alpha \lesssim -0.8$, a wormhole structure with a finite minimum area (corresponding to the throat) arises, whereas for $\alpha \gtrsim -0.8$ the topology remains Euclidean and the solutions represent a kind of spherical condensate.  These results hold true  in both the $\epsilon <0$ and $\epsilon >0$ cases.

Despite having curvature divergences at the center, $r\to 0$, the non-wormhole structure turns out to be geodesically complete due to the fact that the center of the solutions, $r=0$, lies on the future (or past) boundary of the spacetime, as it cannot be reached in finite affine time by geodesic observers or radiation (this is similar to recent results obtained in some Palatini $f(R)$ theories containing wormhole solutions \cite{fRani,fRem}). On the other hand, in the wormhole cases the throat can be reached and crossed by light rays and particles with energy above the maximum of the effective potential barrier. Once crossed, the internal asymptotic infinity is always reached in a finite affine time, putting forward the geodesic incompleteness of these configurations regardless of the behavior of curvature scalars as $y\to \infty$.

An interesting feature of the obtained solutions is related to the total energy density associated to the gravity plus scalar sectors. A solitonic profile has been found representing a lump/well of energy concentrated on a finite region of space around the wormhole throat, while vanishing away from it, which is a typical property of soliton configurations on a flat spacetime. Non-wormhole configurations also exhibit this shell-like distribution of energy, whose maximum is not localized at the center of the solutions. This feature was checked to exist for different values of the Born-Infeld length parameter and different values of the integration constants of the problem.

Remarkably, the location of the wormhole throat is linearly related with the parameter $|\alpha|$, having enough freedom in the choice of parameters and units to make it coincide with the corresponding Schwarzschild radius of a black hole with the same mass, {\it i.e.}, $r_{th}=2M$. This intriguing property arises as a result of the analytical and numerical analysis and by no means could have been anticipated from the field equations before their numerical integration or designed {\it a priori} in an attempt to replace event horizons by wormhole throats.

Whether these objects could have any astrophysical significance is a matter that will be explored in detail elsewhere given the importance of discriminating between black holes and other compact objects, as pointed out recently in the context of gravitational wave emission by Cardoso et al. \cite{GWe,Cardoso:2016rao}, using traversable wormholes (and other horizonless compact candidates) disguised as black holes. In the case of scalar geons considered here, the lack of an event horizon together with the incompleteness of geodesics in the wormhole configurations indicates that they are some kind of naked singularity, but with bounded total energy density (and curvature scalars as well in the case of $\epsilon<0$). On the other hand, the fact that they can be smoothly connected with geodesically complete solutions (those with Euclidean topology) suggests that the wormhole configurations could be unstable or represent transients, and that they might dynamically decay into lighter, more stable forms. The dynamical process of going from Euclidean topology configurations to non-Euclidean ones is likely to represent critical phenomena \cite{Gundlach:2007gc} and is interesting on its own, as is the process of quantum particle creation in those rapidly changing backgrounds \cite{Parker:2009uva, Fabbri:2005mw}. Research in these directions is currently underway.

\section*{Acknowledgments}

V.I.A. is supported by the postdoctoral fellowship CNPq-Brasil/PDE No.~234432/2014-4 and by Federal University of Campina Grande (Brazil). G.J.O. is supported by a Ramon y Cajal contract and the Spanish grant FIS2014-57387-C3-1-P from MINECO. D.R.G. is funded by the Funda\c{c}\~ao para a Ci\^encia e a Tecnologia (FCT, Portugal) postdoctoral fellowship No.~SFRH/BPD/102958/2014 and the FCT research grant UID/FIS/04434/2013. This work has also been supported by the i-COOPB20105 grant of the Spanish Research Council (CSIC), the Consolider Program CPANPHY-1205388, the Severo Ochoa grant SEV-2014-0398 (Spain), and the CNPq (Brazilian agency) project No.301137/2014-5. V.I.A. and D.R.G. thank the Department of Physics of the University of Valencia for their hospitality during the elaboration of this work. This article is based upon work from COST Action CA15117, supported by COST (European Cooperation in Science and Technology).

\section*{Appendix} \label{sec:App1}

In this appendix we provide the main elements justifying the suitability of the choice (\ref{ds}) and (\ref{dst}) for the line elements of the auxiliary and spacetime geometries, respectively. To start with one can introduce two line elements, suitably adapted to the symmetries and the two-metric structure of our problem, as

\begin{eqnarray}
ds^2=-B(x)e^{2\Phi(x)}dt^2+\frac{1}{B(x)}dx^2+r^2(x)d\Omega^2 \ , \label{eq:ds2} \\
d\tilde{s}^2=-\tilde{B}(x)e^{2\tilde{\Phi}(x)}dt^2+\frac{1}{\tilde{B}(x)}dx^2+\tilde{r}^2(x)d\Omega^2 \ , \label{eq:dst2}
\end{eqnarray}
for the spacetime metric $g_{\mu\nu}$ and the auxiliary $q_{\mu\nu}$ metric, respectively. In these line elements, $B(x)$, $\Phi(x)$, $r(x)$, $\tilde{B}(x)$, $\tilde{\Phi}(x)$ and $\tilde{r}(x)$ are functions to be determined through the resolution of the field equations, and which are related to each other via the transformation (\ref{eq:qg}) or, explicitly, by

\begin{equation} \label{eq:relqg}
e^{2\tilde{\Phi}}=\Omega_+\Omega_- e^{2{\Phi}} \hspace{0.1cm};\hspace{0.2cm} B=\Omega_-\tilde{B} \hspace{0.1cm};\hspace{0.2cm} \tilde{r}^2=\Omega_+ r^2 \ .
\end{equation}
With these definitions the first integral of the scalar field equations (\ref{eq:scalarfield}) reads

\begin{equation} \label{eq:scanew}
r^2 e^\Phi B \phi_x=C \ .
\end{equation}
Next we follow Wyman's approach, whose trick lies on employing the scalar field as radial coordinate, {\it i.e.} or, in other words, to make $\phi_x$ to be a constant, $\phi_x=v_0$. Then one can write Eq.(\ref{eq:scanew}) using (\ref{eq:relqg}) as $\tilde{B}=\frac{C_0 e^{-\tilde{\Phi}}}{\tilde{r}^2}\frac{\Omega^{1/2}}{\Omega_-}$, where $C_0=C/v_0$. Inserting this expression into the line element for $q_{\mu\nu}$, Eq.(\ref{eq:dst2}), we obtain

\begin{eqnarray}
d\tilde{s}^2&=&-\left(\tfrac{C_0 e^{\tilde{\Phi}}}{\tilde{r}^2}\tfrac{\Omega^{1/2}}{\Omega_-}\right)dt^2+\left(\tfrac{C_0 e^{\tilde{\Phi}}}{\tilde{r}^2}\tfrac{\Omega^{1/2}}{\Omega_-}\right)\tfrac{\tilde{r}^4}{C_0^2}\left(\tfrac{\Omega_-}{\Omega^{1/2}}dx\right)^2+ \nonumber \\
&&+\tilde{r}^2d\Omega^2 \ . \label{eq:ds2rt}
\end{eqnarray}
The final step is just to use this form of the line element to motivate the introduction of the ansätze given by Eqs.(\ref{ds}) and (\ref{dst}), which simplify many calculations.


\begin{thebibliography}{100}

\bibitem{Wheeler}
J. A. Wheeler, Phys. Rev. \textbf{97}, 511 (1955).
\bibitem{LobosBook}
F. S. N. Lobo (Ed), {\it Wormholes, Warp Drives and Energy Conditions}, Springer Int. Publishing (2017 to appear), doi:10.1007/978-3-319-55182-1.
\bibitem{MW}
C. W. Misner and J. A. Wheeler,  Ann. Phys. \textbf{2}, 525 (1957).
\bibitem{Louko:2004ej}
J.~Louko, R.~B.~Mann and D.~Marolf,
  %``Geons with spin and charge,''
Class.\ Quant.\ Grav.\  {\bf 22}, 1451 (2005).
\bibitem{Olmo:2017qab}
G.~J.~Olmo and D.~Rubiera-Garcia,
  %``Geons in Palatini Theories of Gravity,''
Fundam.\ Theor.\ Phys.\  {\bf 189}, 161 (2017).
\bibitem{Martinon:2017uyo}
  G.~Martinon, G.~Fodor, P.~Grandclément and P.~Forgàcs,
  %``Gravitational geons in asymptotically anti-de Sitter spacetimes,''
Class.\ Quant.\ Grav.\  {\bf 34}, 125012 (2017)
\bibitem{Olmo:2015bha}
G.~J.~Olmo and D.~Rubiera-Garcia,
  %``The quantum, the geon, and the crystal,''
Int.\ J.\ Mod.\ Phys.\ D {\bf 24}, 1542013 (2015).
\bibitem{SolBooks}
R. Rajaraman, \emph{Solitons and Instantons} (North-Holland, 1982);
N. Manton and P. Sutcliffe, \emph{Topological solitons} (Cambridge University Press, 2004);
A. Vilenkin and E. P. S. Shellard, \emph{Cosmic strings and other topological defects} (Cambridge University Press, 1994).
\bibitem{Topo-Sol}
T. H. R. Skyrme, Proc. Roy. Soc. A \textbf{260}, 127 (1961);
H. Nielsen and P. Olesen, Nucl. Phys. B \textbf{61}, 1064 (1973);
R. Jackiw and E. J. Weinberg, Phys. Rev. Lett. \textbf{64}, 2234 (1990);
D.~Bazeia, M.~J.~dos Santos, and R.~F.~Ribeiro, Phys.\ Lett.\ A {\bf 208}, 84 (1995);
E. Babichev, Phys. Rev. D \textbf{74}, 085004 (2006);
\textbf{77}, 065021 (2008).
\bibitem{NTopo-Sol}
R. Friedberg and T. D. Lee, Phys. Rev. D, \textbf{15}, 1694 (1977); \textbf{16}, 1096 (1977);
T. D. Lee and Y. Pang,  Phys. Rept. \textbf{221}, 251 (1992);
S. Coleman, Nucl. Phys. B \textbf{262}, 263 (1985);
D.~Bazeia, L.~Losano, M.~A.~Marques, R.~Menezes, and R.~da Rocha, Phys.\ Lett.\ B {\bf 758}, 146 (2016);
D.~Bazeia, L.~Losano, M.~A.~Marques, and R.~Menezes,   Phys.\ Lett.\ B {\bf 765}, 359 (2017).
\bibitem{TheoSol}
G. H. Derrick, J. Math. Phys \textbf{5}, 1252 (1964);
S. Coleman, Comm. Math. Phys. \textbf{55}, 113 (1977);
S. Coleman and L. Smarr, Comm. Math. Phys. \textbf{56}, 1 (1977).
\bibitem{Non-canonical}
D. Bazeia, E. da Hora, R. Menezes, H. P. de Oliveira, and C. dos Santos, Phys. Rev. D \textbf{81}, 125016 (2010);
C. Adam, J. Sanchez-Guillen, and A. Wereszczynski, Phys. Rev. D \textbf{82}, 085015 (2010);
P. P. Avelino, D. Bazeia, and R. Menezes, Eur. Phys. J. C \textbf{71}, 1683 (2011);
R. Casana, M. M. Ferreira, E. da Hora, and C. dos Santos, Phys. Lett. B \textbf{722}, 193 (2013).
\bibitem{BM}
R. Bartnik and J. McKinnon, Phys. Rev. Lett. \textbf{61}, 141 (1988).
\bibitem{GravSol}
M. S. Volkov and D. V. Gal'tsov, Phys. Rept. \textbf{319}, 1 (1999);
D. Gal'tsov and R. Kerner, Phys. Rev. Lett. \textbf{84}, 5955 (2000);
S. Hod, Phys. Lett. B \textbf{661}, 175 (2008).
\bibitem{Fisher}
I.~Z.~Fisher, Zh.\ Eksp.\ Teor.\ Fiz.\  {\bf 18},  636 (1948) [arXiv:gr-qc/9911008].
\bibitem{HBS}
C. A. R. Herdeiro and E. Radu, Phys. Rev. Lett. \textbf{112}, 221101 (2014);
Class. Quant. Grav. \textbf{32}, 144001 (2015);
G.~Dvali and A.~Gußmann, Nucl.\ Phys.\ B {\bf 913},  1001 (2016).
\bibitem{VolkovReview}
C.~A.~R.~Herdeiro and E.~Radu,  Int.\ J.\ Mod.\ Phys.\ D {\bf 24}, 1542014 (2015);
M.~S.~Volkov, arXiv:1601.08230 [gr-qc].
\bibitem{NH}
W. Israel, Commun. Math. Phys. \textbf{8}, 245 (1968);
B. Carter, Phys. Rev. Lett. \textbf{26}, 331 (1971).
\bibitem{KNhair}
J.~F.~M.~Delgado, C.~A.~R.~Herdeiro, E.~Radu, and H.~Runarsson, Phys.\ Lett.\ B {\bf 761}, 234 (2016);
Y.~Ni, M.~Zhou, A.~Cardenas-Avendano, C.~Bambi, C.~A.~R.~Herdeiro, and E.~Radu, JCAP {\bf 1607}, 049 (2016);
C.~L.~Benone, L.~C.~B.~Crispino, C.~Herdeiro, and E.~Radu, Phys.\ Rev.\ D {\bf 90}, 104024 (2014).
\bibitem{Superradiance}
R. Brito, V. Cardoso, and P. Pani, Lect. Notes Phys. \textbf{906} (2015).
\bibitem{BBB}
V. Cardoso, O. J. C. Dias, J. P. S. Lemos, and S. Yoshida, Phys. Rev. D \textbf{70}, 044039 (2004); 049903 (2004);
V. Cardoso and O. J. C. Dias, Phys. Rev. D \textbf{70}, 084011 (2004);
N.~Sanchis-Gual, J.~C.~Degollado, P.~J.~Montero, J.~A.~Font, and C.~Herdeiro, Phys.\ Rev.\ Lett.\  {\bf 116}, 141101 (2016);
N.~Sanchis-Gual, J.~C.~Degollado, P.~Izquierdo, J.~A.~Font, and P.~J.~Montero,  Phys.\ Rev.\ D {\bf 94}, 043004 (2016); 044061 (2016).
\bibitem{Liebling}
S. L. Liebling and C. Palenzuela, Living Rev. Rel. \textbf{15}, 6 (2012).
\bibitem{GS}
F.~Canfora and H.~Maeda,  Phys.\ Rev.\ D {\bf 87},  084049 (2013);
S.~B.~Gudnason, M.~Nitta, and N.~Sawado, JHEP {\bf 1512},  013 (2015);
E.~Ayon-Beato, F.~Canfora, and J.~Zanelli,  Phys.\ Lett.\ B {\bf 752},  201 (2016).
\bibitem{NSG}
N.~Sanchis-Gual, J.~C.~Degollado, P.~J.~Montero, and J.~A.~Font, Phys.\ Rev.\ D {\bf 91},  043005 (2015);
N.~Sanchis-Gual, J.~C.~Degollado, P.~J.~Montero, J.~A.~Font, and V.~Mewes, Phys.\ Rev.\ D {\bf 92},   083001 (2015);
S.~Ponglertsakul, E.~Winstanley, and S.~R.~Dolan, Phys.\ Rev.\ D {\bf 94},  024031 (2016);
A.~Escorihuela-Tomàs, N.~Sanchis-Gual, J.~C.~Degollado and J.~A.~Font, arXiv:1704.08023 [gr-qc].
\bibitem{Sol-other}
U.~Nucamendi and M.~Salgado, Phys.\ Rev.\ D {\bf 68},  044026 (2003);
C.~Herdeiro, E.~Radu, and H.~Runarsson, Class.\ Quant.\ Grav.\  {\bf 33},    154001 (2016);
S.~Hod, Phys.\ Lett.\ B {\bf 763}, 275 (2016);
E.~Babichev, C.~Charmousis, and M.~Hassaine, JHEP {\bf 1705} 114 (2017);
W.~E.~East and F.~Pretorius, arXiv:1704.04791 [gr-qc].
\bibitem{fs}
V.~Cardoso, S.~Chakrabarti, P.~Pani, E.~Berti, and L.~Gualtieri, Phys.\ Rev.\ Lett.\  {\bf 107},  241101 (2011).
\bibitem{st}
V.~Cardoso, I.~P.~Carucci, P.~Pani, and T.~P.~Sotiriou,  Phys.\ Rev.\ Lett.\  {\bf 111},  111101 (2013);
E.~Berti, V.~Cardoso, L.~Gualtieri, M.~Horbatsch, and U.~Sperhake, Phys.\ Rev.\ D {\bf 87}, 124020 (2013);
Y.~Brihaye and B.~Hartmann, arXiv:1704.04655 [gr-qc].
\bibitem{BIg}
S. Deser and G. W. Gibbons, Class. Quant. Grav. {\bf 15}, L35 (1998).
\bibitem{BIg-app}
M. Ba\~nados, Phys. Rev. D \textbf{77}, 123534 (2008);
M. Ba\~nados, P. G. Ferreira, and C. Skordis, Phys. Rev. D \textbf{79}, 063511 (2009);
P. Pani, V. Cardoso, and T. Delsate, Phys. Rev. Lett. \textbf{107}, 031101 (2011);
P. P. Avelino and R. Z. Ferreira, Phys. Rev. D \textbf{86}, 041501 (2012);
T. Delsate and J. Steinhoff, Phys. Rev. Lett. \textbf{105}, 011101 (2012);
P. Pani, T. Delsate, and V. Cardoso, Phys. Rev. D \textbf{85}, 084020 (2012);
F. Fiorini, Phys. Rev. Lett. \textbf{111}, 011101 (2013);
M.~Bouhmadi-Lopez, C.~Y.~Chen, and P.~Chen, Eur.\ Phys.\ J.\ C {\bf 74},  2802 (2014);
{\bf 75}, 90 (2015);
Phys.\ Rev.\ D {\bf 90},  123518 (2014);
S.~Jana and S.~Kar,  Phys.\ Rev.\ D {\bf 92},  084004 (2015);
R.~Shaikh, Phys.\ Rev.\ D {\bf 92},  024015 (2015);
C.~Y.~Chen, M.~Bouhmadi-Lopez, and P.~Chen, Eur.\ Phys.\ J.\ C {\bf 76},  40 (2016);
P.~P.~Avelino, Phys.\ Rev.\ D {\bf 93},  044067 (2016); 104054 (2016);
S.~Jana and S.~Kar, arXiv:1706.03209 [gr-qc];
K.~Yang, Y.~X.~Liu, B.~Guo and X.~L.~Du, arXiv:1706.04818 [hep-th];
S.~L.~Li and H.~Wei,  arXiv:1705.06819 [gr-qc].
\bibitem{bhor17}
J.~Beltran Jimenez, L.~Heisenberg, G.~J.~Olmo and D.~Rubiera-Garcia, arXiv:1704.03351 [gr-qc].
\bibitem{ors1}
G. J. Olmo, D. Rubiera-Garcia, and A. Sanchez-Puente, Eur. Phys. J. C \textbf{76},  143 (2016).
\bibitem{oor14}
S. D. Odintsov, G. J. Olmo, and D. Rubiera-Garcia, Phys. Rev. D \textbf{90},  044003 (2014).
\bibitem{OlmoRub}
G. J. Olmo, D. Rubiera-Garcia, and H. Sanchis-Alepuz, Eur. Phys. J. C \textbf{74},  2804 (2014).
\bibitem{BF}
M. Ba\~nados and P. G. Ferreira, Phys. Rev. Lett. {\bf 105}, 011101 (2010).
\bibitem{ors15a}
G. J. Olmo, D. Rubiera-Garcia, and A. Sanchez-Puente, Phys. Rev. D \textbf{92}, 044047 (2015).
\bibitem{ors2}
G. J. Olmo, D. Rubiera-Garcia, and A. Sanchez-Puente, Class. Quant. Grav. \textbf{33},  115007 (2016).
\bibitem{Melvin}
M. A. Melvin, Phys. Lett. \textbf{8}, 65 (1964).
\bibitem{Melvin-cor}
C. Bambi, G. J. Olmo, and D. Rubiera-Garcia, Phys. Rev. D \textbf{91},  104010 (2015).
\bibitem{Zanelli}
J. Zanelli, arXiv:hep-th/0502193.
\bibitem{Franca}
M. Ferraris, M. Francaviglia, and I. Volovich, Class. Quant. Grav. \textbf{11}, 1505 (1994);
A. Borowiec, M. Ferraris, M. Francaviglia, and I. Volovich, Class. Quant. Grav. \textbf{15}, 43 (1998).
\bibitem{ReviewsPal}
S. Capozziello and M. De Laurentis, Phys. Rep. \textbf{509}, 167 (2011);
S. Nojiri and S. D. Odintsov, Phys. Rep. \textbf{505}, 59 (2011).
\bibitem{GWe}
V.~Cardoso, S.~Hopper, C.~F.~B.~Macedo, C.~Palenzuela, and P.~Pani,  Phys.\ Rev.\ D {\bf 94},  084031 (2016).
\bibitem{GWeb}
J.~Abedi, H.~Dykaar, and N.~Afshordi, arXiv:1612.00266 [gr-qc];
C.~Barcel\'o, R.~Carballo-Rubio and L.~J.~Garay, JHEP {\bf 1705} 054 (2017).
\bibitem{LIGO}
B.~P.~Abbott {\it et al.} [LIGO Scientific and Virgo Collaborations], Phys.\ Rev.\ X {\bf 6}, 041015 (2016).
\bibitem{Lobo:2013prg}
F.~S.~N.~Lobo, G.~J.~Olmo and D.~Rubiera-Garcia, JCAP {\bf 1307}, 011 (2013);
J.~Khoury, Phys.\ Rev.\ D {\bf 91}, 024022 (2015);
S.~Clesse and J.~García-Bellido, Phys.\ Rev.\ D {\bf 92}, 023524 (2015);
J.~Garcia-Bellido and E.~Ruiz Morales, arXiv:1702.03901 [astro-ph.CO];
J.~García-Bellido, arXiv:1702.08275 [astro-ph.CO].
\bibitem{or13}
G. J. Olmo and D. Rubiera-Garcia, JCAP \textbf{1402}, 010 (2014).
\bibitem{Torsion}
G. J. Olmo and D. Rubiera-Garcia, Phys. Rev. D \textbf{88}, 084030 (2013).
\bibitem{Jimenez:2015caa}
J.~Beltran Jimenez, L.~Heisenberg, and G.~J.~Olmo, JCAP {\bf 1506}, 026 (2015).
\bibitem{Bazeia:2015zpa}
D.~Bazeia, L.~Losano, R.~Menezes, G.~J.~Olmo, and D.~Rubiera-Garcia, Class.\ Quant.\ Grav.\  {\bf 32},  215011 (2015).
\bibitem{Wyman}
M. Wyman, Phys. Rev. D \textbf{24}, 839 (1981).
\bibitem{Janis:1968zz}
  A.~I.~Janis, E.~T.~Newman and J.~Winicour,
  %``Reality of the Schwarzschild Singularity,''
  Phys.\ Rev.\ Lett.\  {\bf 20}, 878 (1968).
\bibitem{Visser}
M. Visser, \emph{Lorentzian wormholes} (Springer-Verlag, New York, 1995).
\bibitem{Theorems}
R. Penrose, Phys. Rev. Lett. \textbf{14}, 57 (1965);
Riv. Nuovo Cim. Numero Speciale \textbf{1}, 252 (1969);
Gen. Relativ. Gravit. \textbf{34}, 1141 (2002);
S. W. Hawking, Phys. Rev. Lett. \textbf{17}, 444 (1966).
\bibitem{SG15}
J.~M.~M.~Senovilla and D.~Garfinkle, Class.\ Quant.\ Grav.\  {\bf 32}, 124008 (2015).
\bibitem{LecOlmo}
G.~J.~Olmo, Springer Proc.\ Phys.\  {\bf 176},  183 (2016) [arXiv:1607.06670 [hep-th]].
\bibitem{Wald}
R. M. Wald, \emph{General Relativity} (University Press, Chicago, 1984)
\bibitem{Chandra}
S. Chandrasekhar, \emph{The Mathematical Theory of Black Holes} (Oxford University Press, New York, 1992).
\bibitem{fRani}
G. J. Olmo and D. Rubiera-Garcia, Universe \textbf{2015}, 173;
C.~Bejarano, G.~J.~Olmo, and D.~Rubiera-Garcia, Phys.\ Rev.\ D {\bf 95}, 064043 (2017).
\bibitem{fRem}
C. Bambi, A. Cardenas-Avendano, G. J. Olmo, and D. Rubiera-Garcia, Phys. Rev. D \textbf{93}, 064016 (2016).
\bibitem{Cardoso:2016rao}
V.~Cardoso, E.~Franzin, and P.~Pani, Phys.\ Rev.\ Lett.\  {\bf 116}, 171101 (2016); Erratum: Phys.\ Rev.\ Lett.\  {\bf 117}, 089902 (2016).
\bibitem{Gundlach:2007gc}
C.~Gundlach and J.~M.~Martin-Garcia, Living Rev.\ Rel.\  {\bf 10}, 5 (2007).
\bibitem{Parker:2009uva}
L.~E.~Parker and D.~Toms, \emph{Quantum Field Theory in Curved Spacetime: Quantized Field and Gravity} (Cambridge University Press, 2009).
\bibitem{Fabbri:2005mw}
A.~Fabbri and J.~Navarro-Salas, \emph{Modeling black hole evaporation} (Imp. Coll. Pr., London, UK, 2005).

\end{thebibliography}
\end{document}